\documentclass[]{jkas} 

\def\year{2024} 
\def\volume{56} 
\def\issue{1} 
\def\beginpage{1} 
\def\received{---} 
\def\accepted{---} 
\def\published{---} 
\date{Received \received; Accepted \accepted; Published \published}

\newcommand{\refbf}{}

\title{%
KMTNet Synoptic Survey of Southern Sky III: The First Data Release
}

\author[1]{Seo-Won Chang}{0000-0002-3118-8275}
\author[1$\star$]{Myungshin Im}{0000-0002-8537-6714}
\author[1]{Mankeun Jeong}{0009-0003-1280-0099}
\author[2]{Joonho Kim}{0000-0002-1294-168X}
\author[1]{Bomi Park}{0009-0008-3499-3043}
\author[1]{Jaewon Lee}{0009-0005-3910-0337}
\author[3,4,5]{David A. H. Buckley}{0000-0002-7004-9956}
\author[6,7]{Jeff Cooke}{0000-0001-5703-2108}
\author[1]{Sungho Jung}{0009-0009-5870-4266}
\author[8]{Dong-Jin Kim}{0000-0002-4292-9649}
\author[1]{Ji Hoon Kim}{0000-0002-1418-3309}
\author[9,10]{Yongjung Kim}{0000-0003-1647-3286}
\author[8]{Chung-Uk Lee}{0000-0003-0043-3925}
\author[1]{Seong-Kook Lee}{0000-0001-5342-8906}
\author[11]{Gregory S. H. Paek}{0000-0002-6639-6533}
\author[1]{Jiseop Shin}{0009-0005-3944-1457}

\affil[1]{SNU Astronomy Research Center, Astronomy Program, Department of Physics and Astronomy, Seoul National University, Gwanak-gu, Seoul 08826, Republic of Korea}
\affil[2]{Daegu National Science Museum, 20, Techno-daero 6-gil, Yuga-myeon, Dalseong-gun, Daegu 43023, Republic of Korea}
\affil[3]{Department of Astronomy, University of Cape Town, Private Bag X3, Rondebosch 7701, South Africa}
\affil[4]{South African Astronomical Observatory, P.O Box 9, Observatory, 7935 Cape Town, South Africa}
\affil[5]{Department of Physics, University of the Free State, PO Box 339, Bloemfontein 9300, South Africa}
\affil[6]{Centre for Astrophysics and Supercomputing, Swinburne University of Technology, John Street, Hawthorn, VIC 3122, Australia}
\affil[7]{ARC Centre of Excellence for Gravitational Wave Discovery (OzGrav), John Street, Hawthorn, VIC 3122, Australia}
\affil[8]{Korea Astronomy and Space Science Institute, Daejeon 34055, Republic of Korea}
\affil[9]{School of Liberal Studies, Sejong University, 209 Neungdong-ro, Gwangjin-Gu, Seoul 05006, Republic of Korea}
\affil[10]{Department of Physics and Astronomy, Sejong University, 209 Neungdong-ro, Gwangjin-Gu, Seoul 05006, Republic of Korea} 
\affil[11]{Institute for Astronomy, University of Hawaii, 2680 Woodlawn Drive, Honolulu, HI 96822, USA}

\def\corrauthor{%
M. Im, \email{myungshin.im@gmail.com}; S.-W. Chang, \email{seowon.chang@snu.ac.kr} (lead)

}

\def\runningauthor{%
Chang et al.
}

\def\runningtitle{%
KS4 Data Release 1
}

\def\keywords{%
optical astronomy --- surveys --- catalogs --- astronomical databases
}

\def\abstracttext{
We present the first public data release (DR1) of the KMTNet Synoptic Survey of Southern Sky (KS4). This deep, wide-field imaging survey covers a southern footprint of -85$^{\circ}$ < Decl. < -28.8$^{\circ}$ in the $B$, $V$, $R$, and $I$ bands using a network of three 1.6-m telescopes. Although primarily designed to secure reference imaging for gravitational wave counterpart identification, DR1 delivers science-ready data for $\sim$4,000 deg$^{2}$ to enable a broad range of astrophysical research. The release includes deep co-added images reaching median 5$\sigma$ depths of 22.0-23.5 AB mag. It is accompanied by two source catalogs containing over 200 million sources with SNR $>5$: an $I$-band-selected forced-photometry catalog optimized for consistent colors, and a band-merged catalog offering enhanced completeness. Validation demonstrates robust data quality, characterized by mean astrometric offsets of $+0.054 \pm 0.129$ arcsec in RA and $-0.015 \pm 0.120$ arcsec in Dec relative to Gaia DR3. {\refbf Photometric uniformity for point sources is maintained within $\pm 0.03$ mag relative to Gaia XP for 97.5--99.8\% of the footprint across all four bands.} A key advantage of KS4 is its uniform and contiguous spatial coverage. It extends to fainter magnitudes than other uniform surveys while filling irregular gaps in existing deep datasets. All data products are publicly available via the CDS and NOIRLab's Astro Data Lab.
}

\begin{document}
\jkashead 


\section{Introduction}
\label{sec:introduction}

Ground-based wide-field optical surveys in the southern hemisphere are increasingly driving open science. {\refbf Recent years have seen the public release of comprehensive datasets---broadband photometric catalogs and imaging data---covering extensive sky areas and hundreds of millions to billions of celestial objects.} Observational strategies vary widely, ranging from comprehensive but relatively shallow all-sky coverage, such as SkyMapper DR4 (\citealt{2024PASA...41...61O}; reaching $\sim$20--22 mag), to deeper unified surveys (e.g., DECam Local Volume Exploration Survey DR2: \citealt{2022ApJS..261...38D}; NOIRLab Source Catalog DR2: \citealt{2021AJ....161..192N}; Legacy Survey DR10: \citealt{2019AJ....157..168D}; reaching $\sim$22.5--24.5 mag). The latter group typically involves the systematic reprocessing of multi-program observations through unified reduction pipelines. The landscape of southern sky surveys will be further transformed by the Vera C. Rubin Observatory's upcoming Legacy Survey of Space and Time (LSST). LSST promises to deliver ultra-deep imaging (down to $\sim$25--27.5 mag) from its 10-year stacks, alongside shallower all-sky coverage ($\sim$23.5--24.5 mag) with a cadence of 3–4 nights (\citealt{2019ApJ...873..111I}). This paradigm of large-scale {\refbf surveys with public data releases}, now central to modern astronomy, builds upon the scientific legacy established by the Sloan Digital Sky Survey (SDSS; \citealt{2000AJ....120.1579Y}). SDSS demonstrated the transformative power of publicly available data, leading to numerous serendipitous discoveries. Examples include new dwarf galaxies (e.g., \citealt{2006ApJ...643L.103Z, 2007ApJ...654..897B}), hypervelocity stars (e.g., \citealt{2005MNRAS.363..223G, 2006ApJ...640L..35B}), and Green Pea galaxies (e.g., \citealt{2009MNRAS.399.1191C}). Furthermore, the data served as a foundation for new methodologies, including applications of machine learning (e.g.,\citealt{2004MNRAS.348.1038B, 2004PASP..116..345C}).



In the context of these expanding data resources, the KMTNet Synoptic Survey of Southern Sky (KS4; Im et al., in preparation) presents its first public data release (DR1). KS4 utilizes the Korea Microlensing Telescope Network (KMTNet; \citealt{Kim2016JKAS...49...37K}), comprising three identical 1.6-m telescopes located at the Cerro Tololo Inter-American Observatory (CTIO) in Chile, the South African Astronomical Observatory (SAAO) in South Africa, and the Siding Spring Observatory (SSO) in Australia. Observations commenced at SAAO on November 29, 2019, initially operating from a single site. Operations subsequently expanded to the full three-site network from October 11, 2020, through September 22, 2025. An early data release (EDR), covering approximately 2,500 deg$^{2}$ (Decl. $< -30^{\circ}$), was previously available to internal collaborators. 

The current DR1 improves significantly upon the EDR. It increases survey coverage by a factor of 1.6 and incorporates enhanced image processing and photometric calibration procedures. While KS4 was primarily designed to secure deep reference imaging for the rapid identification of optical counterparts to gravitational wave events (e.g., \citealt{2021ApJ...916...47K}), the EDR has already demonstrated broader scientific utility. For instance, this early dataset enabled the identification of 72,964 unobscured low-redshift quasar candidates, achieving an 87\% recovery rate for spectroscopically confirmed quasars at z $ <$ 2 (\citealt{2024ApJS..275...46K}). The improved quality of the DR1 dataset now supports direct extragalactic research, driving ongoing searches for high-redshift (z$\sim$5.5), and obscured quasar candidates, as well as galaxy clusters (Park et al., in preparation).

The KMTNet operational team provides standard pre-processing (e.g., bias subtraction and flat-fielding). However, achieving science-ready quality requires further mitigation of instrumental systematics. To address effects not handled by the standard pipeline, we developed enhanced reduction algorithms for KS4 DR1 (see \citealt{2026arXiv2603.17442J} for details). Key improvements for single-epoch image processing include: (i) correction for bright-star-induced bleeding, (ii) residual crosstalk mitigation, (iii) enhanced masking of cosmic rays and static CCD defects, and (iv) photometric zero-point (ZP) homogenization across the survey footprint. {\refbf For detailed technical descriptions and quantitative uncertainty assessments (e.g., astrometric solutions, photometric calibration, and other reduction procedures), we refer the reader to our companion pipeline paper \citep{2026arXiv2603.17442J} for a complete understanding of the data reduction process.} A distinct feature of DR1 is the implementation of multi-observatory image stacking to maximize coverage efficiency. This strategy has proven effective for transient follow-up targeting optical counterparts of high-energy phenomena, including gamma-ray bursts \citep{2024Natur.626..742Y} and gravitational wave events (e.g., \citealt{2024ApJ...960..113P, 2025ApJ...981...38P, 2025arXiv250315422P}). Furthermore, these data products will serve as a critical resource for transient follow-up studies in the era of the Rubin LSST. {\refbf We note that DR1 focuses on delivering deep co-added images 
and photometric catalogs; time-domain data products utilizing the multi-epoch observations are planned for future releases.} 

{\refbf All data products are publicly available to the community. The photometric catalogs are hosted at the Astro Data Lab \citep{2014SPIE.9149E..1TF, 2020A&C....3300411N}, providing access to $BVRI$ photometry for over 200 million sources via standard database query tools. Imaging data are distributed through the Centre de Données astronomiques de Strasbourg (CDS) in the Hierarchical Progressive Survey \citep[HiPS;][]{2015A&A...578A.114F} format, enabling interactive visualization and on-demand cutout generation. Supplementary data products, including photometric zero-point correction maps and quality assessment flags, are also provided.}


The paper is organized as follows. Section \ref{sec:observations and data collection} provides an overview of the KMTNet facilities and the observational strategy. The data reduction and post-processing pipelines are detailed in Section \ref{sec:data reduction and post-processing}. We describe the data products, including images and source catalogs, in Section \ref{sec:data products}, followed by a comprehensive validation of data quality in Section \ref{sec:data validation and quality assessment}. Instructions for accessing the data are provided in Section \ref{sec:data access}. Finally, Section \ref{sec:summary} summarizes the work and discusses future prospects.

\begin{table}
\centering
\caption{DR1 observational summary for 979 KS4 tiles}
\label{tab:dr1_stats}
\begin{tabular}{lcccccc}
\hline
 & \multicolumn{3}{c}{KMTNet Site} & & \\
\cline{2-4}
Filter & CTIO & SSO & SAAO & Total & Exp. Time \\
       &      &     &      & Images & (hours) \\
\hline
$B$ & 1,560 & 762 & 1,925 & 4,247 & 142.0 \\
$V$ & 1,456 & 850 & 1,988 & 4,294 & 143.8 \\
$R$ & 1,416 & 899 & 2,260 & 4,575 & 153.1 \\
$I$ & 1,138 & 623 & 2,749 & 4,510 & 150.4 \\
\hline
Total & 5,570 & 3,134 & 8,922 & {\refbf 17,626} & {\refbf 589.3} \\
\hline
\end{tabular}
\end{table}

\section{Observations and Data Collection}
\label{sec:observations and data collection}
The DR1 dataset includes observations acquired from the commencement of KS4 operations through December 21, 2023. It covers 979 distinct survey tiles, representing a subset of the 2,748 total fields defined in the survey strategy (see Im et al., in preparation). Each tile corresponds to the KMTNet field-of-view ($2 \times 2$~deg$^{2}$) and is typically observed with a 120\,s exposure in each filter, yielding a cumulative integration time of at least 480\,s across the four $BVRI$ bands. Table~\ref{tab:dr1_stats} presents the observational statistics. The dataset includes 17,626 individual exposures obtained across the three KMTNet sites. The contribution from each site varies significantly: SAAO provided 50.5\% of the total exposures (8,922 images), followed by CTIO with 31.5\% (5,570 images) and SSO with 17.7\% (3,134 images). This imbalance primarily reflects differences in site-specific weather conditions, with the SAAO fraction further augmented by supplementary telescope time secured through collaborative programs. The total integration time amounts to 589.3 hours, distributed approximately uniformly across the photometric bands: $B$ (142.0 hr), $V$ (143.8 hr), $R$ (153.1 hr), and $I$ (150.4 hr).

The KMTNet focal plane features a $2\times2$ mosaic of four CCDs, creating physical gaps between adjacent chips. To fill these inter-chip gaps, we adopted a four-point dithering strategy with offsets of 4 arcmin in Right Ascension (RA) and 7 arcmin in Declination (Dec) relative to the tile center. While this approach ensures contiguous spatial coverage, it inevitably introduces systematic variations in imaging depth across the tile. Ideally, the survey strategy called for acquiring all four dither positions at a single site under uniform conditions. However, strict quality control criteria necessitated the rejection of exposures suffering from poor seeing or insufficient depth. Hence, replacement observations were often acquired from alternative sites or at widely separated epochs, resulting in deviations from the nominal strategy. Table~\ref{tab:exposure_distribution} summarizes the resulting distribution of exposure counts per tile. This inhomogeneity stems from the complex interplay between quality control cuts and multi-site scheduling, where manual overrides occasionally produced redundant observations. Approximately 75\% of filter-tile combinations met the baseline requirement of four exposures, while the remaining 25\% contain additional frames from repeated sampling. This observational heterogeneity results in variable photometric depth across the DR1 dataset.

\begin{table*}
\centering
\caption{Distribution of Individual Exposure Counts per Tile}
\label{tab:exposure_distribution}
\begin{tabular}{lcccccc}
\hline\hline
 & \multicolumn{6}{c}{$N_{\rm exp}$} \\ 
\cline{2-7}
Filter & 4 & 5 & 6 & 7 & 8 & $\geq$9 \\
\hline
$B$ & 786 (80.3\%) & 114 (11.6\%) & 44 (4.5\%) & 18 (1.8\%) & 6 (0.6\%) & 11 (1.1\%) \\
$V$ & 761 (77.7\%) & 126 (12.9\%) & 55 (5.6\%) & 14 (1.4\%) & 9 (0.9\%) & 14 (1.4\%) \\
$R$ & 643 (65.7\%) & 155 (15.8\%) & 98 (10.0\%) & 44 (4.5\%) & 23 (2.3\%) & 16 (1.6\%) \\
$I$ & 736 (75.2\%) & 103 (10.5\%) & 46 (4.7\%) & 21 (2.1\%) & 49 (5.0\%) & 24 (2.5\%) \\
\hline
Overall & 2926 (74.7\%) & 498 (12.7\%) & 243 (6.2\%) & 97 (2.5\%) & 87 (2.2\%) & 65 (1.7\%) \\
\hline\hline
\end{tabular}
\vspace{0.9em}
\begin{minipage}{0.75\textwidth}
\small
\textit{Note.} $N_{\exp}$ denotes the number of individual 120\,s exposures obtained for a given tile. The main values represent the count of tiles for each $N_{\exp}$ bin, while percentages in parentheses indicate the fraction relative to the total 979 tiles for that filter.
\end{minipage}
\end{table*}

\begin{figure}
    \centering
    \includegraphics[width=\linewidth]{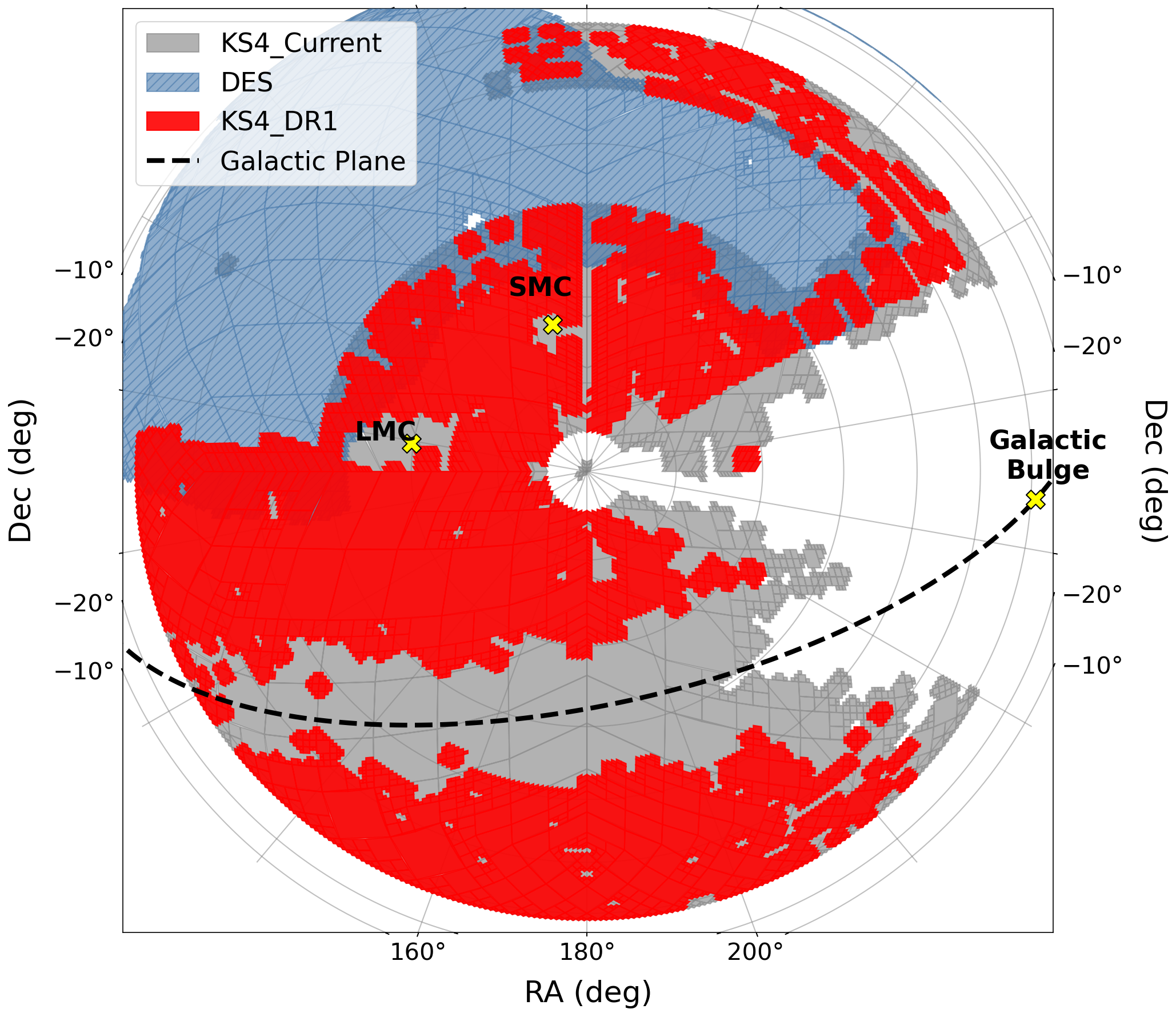}
    \caption{Multi-Order Coverage map of the KS4 DR1 sky coverage, presented in a polar projection centered on the South Celestial Pole (Decl. = -90$^\circ$). The grid lines are spaced at 20-degree intervals in RA and 10-degree intervals in Dec. The red-shaded regions indicate the DR1 footprint. The gray-shaded regions represent the full KS4 survey area observed to date, while the blue hatching shows the extent of the Dark Energy Survey footprint. The positions of the LMC, SMC, and the Galactic Bulge are marked with yellow symbols, and a thick dashed line indicates the Galactic Plane.} 
    \label{fig:coverage_map}
\end{figure}

Figure~\ref{fig:coverage_map} illustrates the DR1 sky coverage using Multi-Order Coverage (MOC) maps, which accurately capture both the irregular survey boundaries and the overlaps between adjacent tiles. This polar projection skymap shows the DR1 footprint (red), the full KS4 survey area observed to date (gray), with the Dark Energy Survey (DES: \citealt{2021ApJS..255...20A}) DR2 footprint (light blue hatching) overplotted for context. The DR1 footprint spans 4,044 deg$^{2}$ in the southern hemisphere, representing 60.1\% of the total area observed to date (6,732 deg$^{2}$). The latter encompasses all acquired observations, including data reserved for future releases. The survey strategy explicitly excluded the Large and Small Magellanic Clouds and minimized overlap with the DES footprint. Despite these exclusions, a substantial overlap of 621 deg$^{2}$ (15.4\% of DR1) with the DES area remains, offering valuable opportunities for photometric cross-validation. Although low Galactic latitude regions were initially planned for exclusion, the current dataset includes coverage of the Galactic plane. Regions with Decl. < -85$^{\circ}$ are excluded due to the mechanical limits of the KMTNet facilities, while the Galactic bulge remains largely unsampled due to competing observational priorities with the microlensing survey.

\section{Data Reduction and Post-Processing}
\label{sec:data reduction and post-processing}
A comprehensive description of the data processing pipeline developed for KS4 DR1 is provided in the companion paper by \citep{2026arXiv2603.17442J}. In this section, we outline the key processing steps and calibration procedures utilized to generate the deep co-added images and the associated photometric catalogs.

\subsection{Pipeline Overview} 
\label{sec:pipeline overview}

\subsubsection{Single-epoch images}
\label{sec:single-epoch images}
Raw data from the three KMTNet sites are transferred to the data processing center in Daejeon, Korea, for initial reduction. Standard pre-processing procedures are applied, including overscan and crosstalk corrections, bias subtraction, and flat-fielding using master twilight flats. Following this, the reduced images are processed via the dedicated KS4 pipeline, which handles mask generation, source detection, and both astrometric and photometric calibration.

KMTNet images are affected by various detector artifacts that can severely degrade source detection and photometric accuracy. To address this, we generate bad pixel masks by automatically identifying contaminated regions within individual exposures. The masking algorithm identifies cosmic rays {\refbf (via the L.A.Cosmic algorithm; \citealt{2001PASP..113.1420V})}, crosstalk artifacts arising from saturated sources \citep{Kim2016JKAS...49...37K}, bleeding trails (Shin et al., in preparation), malfunctioning readout amplifier segments, and intrinsic CCD defects. These features are encoded into a bitmask using specific flags: 1 (cosmic rays and crosstalk), 2 (bleeding), 4 (CCD defects), and 8 (amplifier defects). {\refbf Further technical details on each masking procedure are provided in Section 3.3 of \citep{2026arXiv2603.17442J}.


The pre-processed data are scaled to a common reference ZP of 30.0 mag. This correction is applied independently to each readout amplifier to correct for ZP variations primarily along the Y-axis (Dec). {\refbf Prior to correction, ZP variations typically reach $\sim$0.1 mag within amplifiers and $\sim$0.2 mag between sites. Their consistency across filters suggests a common instrumental origin (see Figures 4 and 12 in \citealt{2026arXiv2603.17442J}). Variations along the X-axis (RA) within individual amplifiers were found to be negligible. Nevertheless, residual vertical striping along amplifier boundaries persists in a subset of images, particularly in the $I$-band. These artifacts originate from imperfect flat-fielding rather than the ZP scaling process itself and cannot be removed even with 2D ZP fitting. To assist users in identifying and excluding these problematic regions, a list of affected tile and filter combinations is provided in the public repository (see Section 5.4.1 in \citealt{2026arXiv2603.17442J}).


Astrometric solutions are derived using SCAMP \citep{2006ASPC..351..112B}, incorporating pre-computed focal plane distortion models (stored as global ahead files) as priors. These solutions are validated against Gaia EDR3 \citep{2016A&A...595A...1G,2021A&A...649A...1G}, which serves as the primary reference catalog. To ensure robust alignment across the focal plane, each CCD chip is partitioned into 64 sub-sections. We enforce a strict quality criterion requiring that at least 62 of these sub-sections demonstrate a source match rate $> 60\%$ with Gaia and root-mean-square (RMS) alignment errors $< 0.5$ arcsec. This procedure yields a typical astrometric precision of 0.026 arcsec in both RA and Dec.

\subsubsection{Co-added images}
\label{sec:co-added images}
Following astrometric calibration, individual exposures are co-added using SWarp \citep{2010ascl.soft10068B} to generate deep stacked images. The stacking procedure adopts a tangential projection with LANCZOS3 resampling, incorporating a median combination method to robustly reject artifacts. We apply fixed center coordinates to ensure that all bands share a common pixel grid, thereby enabling consistent multi-band photometry and difference image analysis without the need for further resampling. The final co-added images exhibit median seeing values of 2.10, 2.03, 1.96, and 1.93 arcsec for the $B$, $V$, $R$, and $I$ filters, respectively. The corresponding median 5$\sigma$ limiting magnitudes are 22.72, 22.56, 22.75, and 22.07 AB mag. While the precise derivation and statistical distributions of these quality metrics are detailed in \citep{2026arXiv2603.17442J}, their spatial distributions across the footprint are presented in this work as skymaps (Fig. \ref{fig:KS4 DR1 seeing} and \ref{fig:KS4 DR1 depth}).

\begin{figure}[!h]
    \centering
    \includegraphics[width=\linewidth]{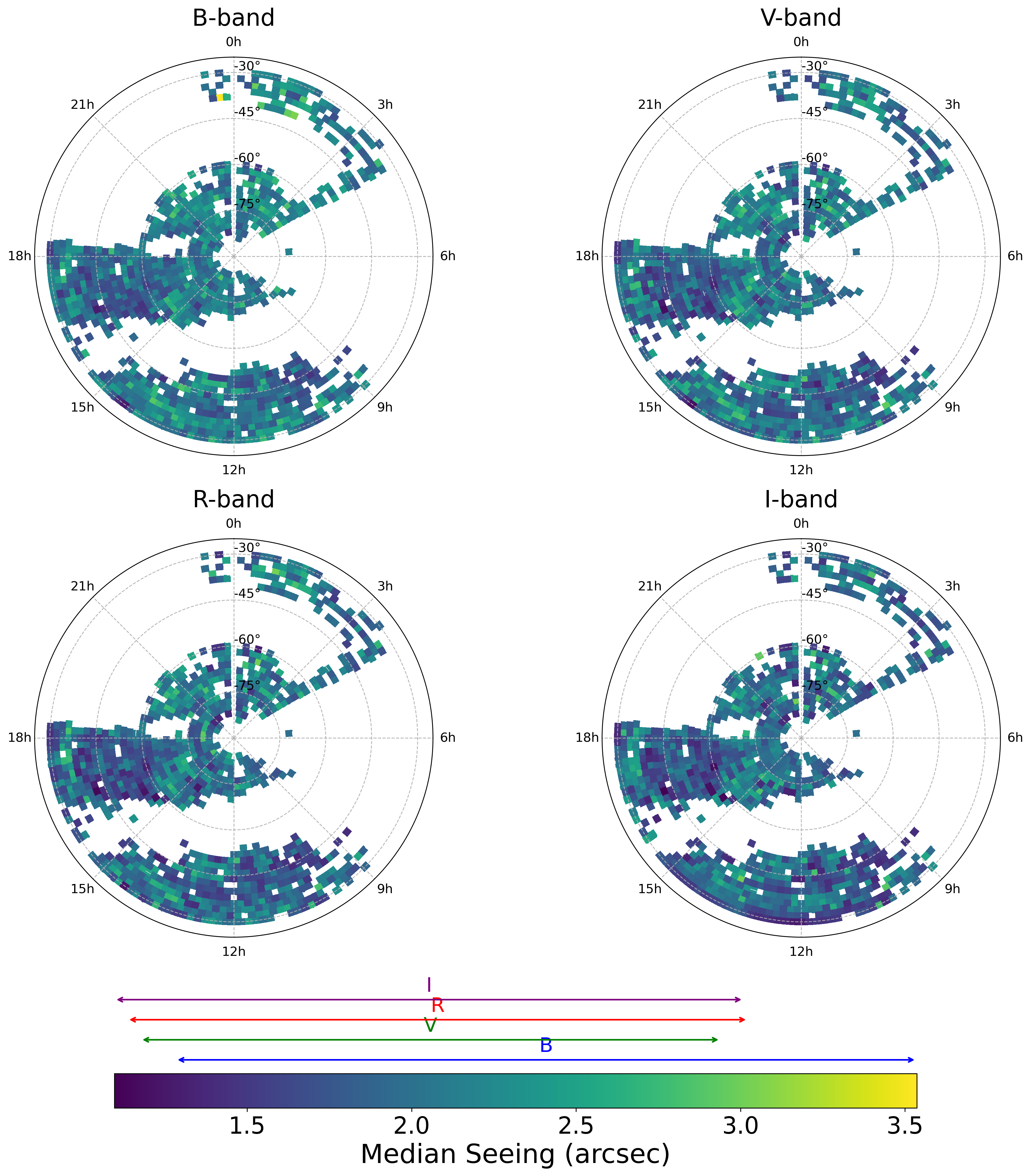}
    \caption{Spatial distribution of seeing (FWHM) for the KS4 DR1 co-added images. Values represent the median FWHM per tile, prioritizing the minimum seeing in overlap regions. The color scale is fixed across all panels for comparison. The full observed ranges are: $B$ (1.28--3.54 arcsec), $V$ (1.18--2.94 arcsec), $R$ (1.14--3.02 arcsec), and $I$ (1.10--3.01 arcsec).}
    \label{fig:KS4 DR1 seeing}
\end{figure}

\begin{figure}[!h]
    \centering
    \includegraphics[width=\linewidth]{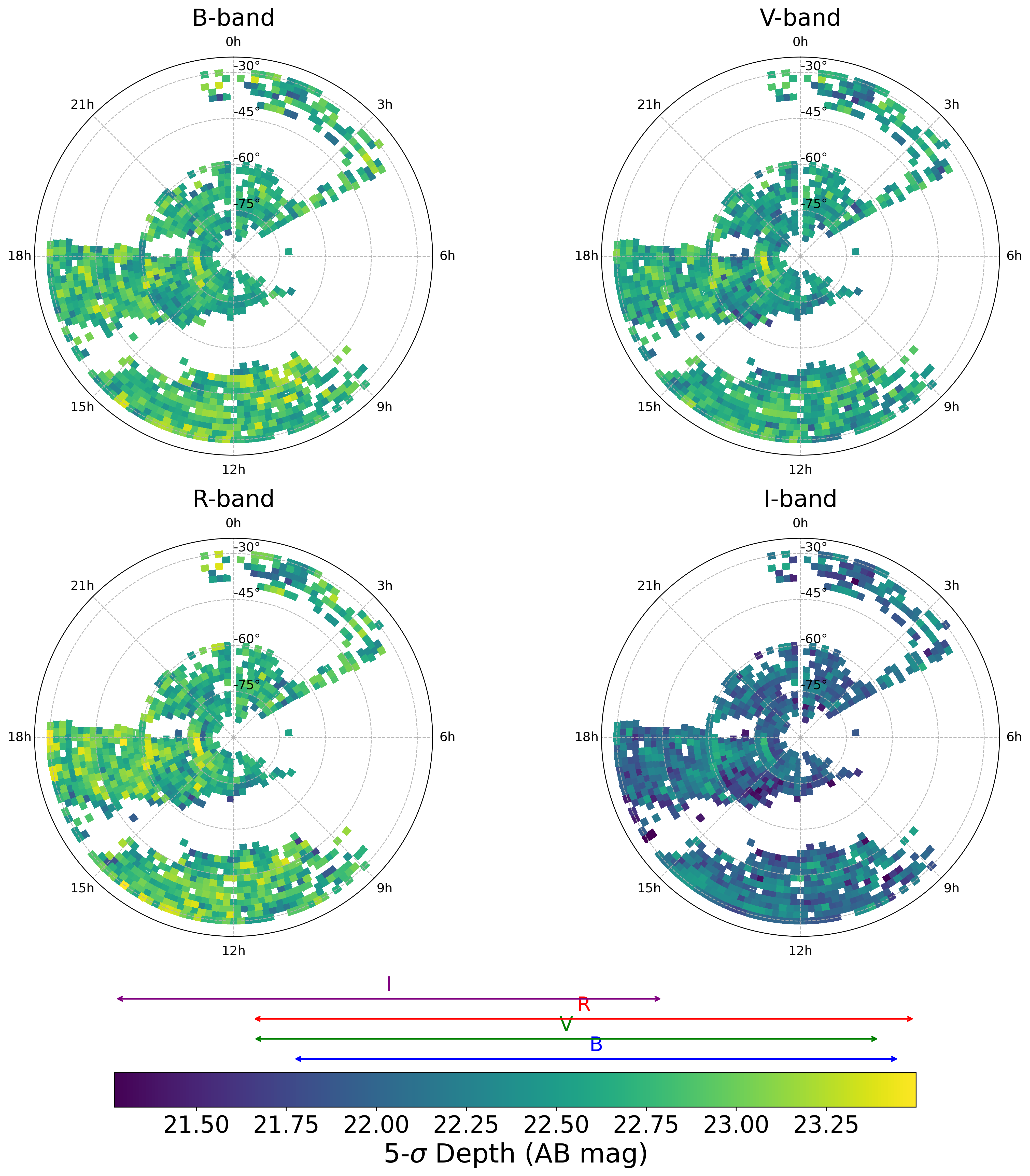}
    \caption{Spatial distribution of $5\sigma$ limiting magnitudes (AB) for the KS4 DR1 co-added images. Values represent the median depth per tile, prioritizing the maximum depth in overlap regions. The color scale is fixed across all panels, with range markers indicating the full extent for each filter. The observed ranges are: $B$ (21.77--23.45 mag), $V$ (21.66--23.40 mag), $R$ (21.65--23.62 mag), and $I$ (21.10--22.80 mag).}    \label{fig:KS4 DR1 depth}
\end{figure}

To ensure photometric uniformity across the data products, we implement a two-stage calibration strategy. First, prior to co-addition, each single-epoch image is subject to an initial ZP standardization. This step is critical for mitigating photometric offsets arising from observations acquired at different epochs and sites. For this preliminary phase, we adopt transformed magnitudes from APASS DR9 (\citealt{2016yCat.2336....0H} for $BVR$) and SkyMapper DR3 (\citealt{2019PASA...36...33O} for $I$). Further details on this initial calibration are provided in \citep{2026arXiv2603.17442J}. However, the co-addition of images obtained under varying seeing conditions inevitably results in residual ZP spatial variations in the stacked frame. Therefore, we applied a second, high-precision calibration to the final co-added images using the Gaia XP synthetic photometry catalog \citep{2023A&A...674A..33G} as a reference. This procedure derives a two-dimensional correction map to model and correct for these spatial non-uniformities. The synthetic photometry uses Johnson-Cousins filter definitions with central wavelengths from \citet{2012PASP..124..140B}. As discussed in Section \ref{sec:known limitations}, while these standard passbands do not perfectly match the KMTNet filter transmission curves, Gaia XP remains the most robust reference for this calibration. This process anchors the final $BVRI$ photometry to the AB system. The DR1 stacked images are flux-scaled to achieve uniform ZPs within 5 arcsec apertures; users may apply the provided local ZP correction maps for other apertures (see Section~\ref{sec:zp map access}).

\subsubsection{Source Detection}
\label{sec:source detection}
We perform source detection using SExtractor \citep{1996A&AS..117..393B} in two distinct modes, applying identical detection settings to optimize source recovery and minimize losses due to image artifacts. First, we run SExtractor in dual-image mode, utilizing the $I$-band as the detection reference while performing forced photometry at the corresponding positions in the $B$, $V$, and $R$ bands. This strategy is specifically designed for high-redshift quasar and galaxy cluster searches, ensuring consistent color measurements for faint sources across all passbands. Second, we conduct independent source detection in each $BVRI$ filter to generate a general-purpose catalog. Both modes provide photometry using 3, 5, and 10 arcsec fixed circular apertures (\texttt{MAG\_APER3}, \texttt{MAG\_APER5}, \texttt{MAG\_APER10}), alongside an adaptively scaled aperture (\texttt{MAG\_AUTO}) designed to measure the total flux of both point and extended sources.

To optimize the detection process, we configure several key SExtractor parameters based on the pipeline described in \citep{2026arXiv2603.17442J}. We use a detection threshold (\texttt{DETECT\_THRESH}) of $1.0\sigma$ above the local background, coupled with a minimum area requirement of 9 contiguous pixels (\texttt{DETECT\_MINAREA}). To minimize the impact of saturated stars, we utilize external masks applied via the \texttt{MASK\_TYPE = CORRECT} setting to interpolate over contaminated regions. This procedure prevents the loss of real sources masked by bleed trails and effectively suppresses spurious detections arising from inaccurate local background estimation.

We set the minimum contrast parameter for de-blending (\texttt{DEBLEND\_MINCONT}) to 0.00005 to improve the separation of closely blended sources. Furthermore, we configure the memory pixel stack (\texttt{MEMORY\_PIXSTACK}) to $10^{7}$ pixels to accommodate the processing of large-format images without buffer overflow. Additionally, the \texttt{IMAFLAGS\_ISO} parameter tracks image artifacts for individual sources by propagating quality flags from the original single-epoch images.

\begin{figure*}
    \centering
    \includegraphics[width=\linewidth]{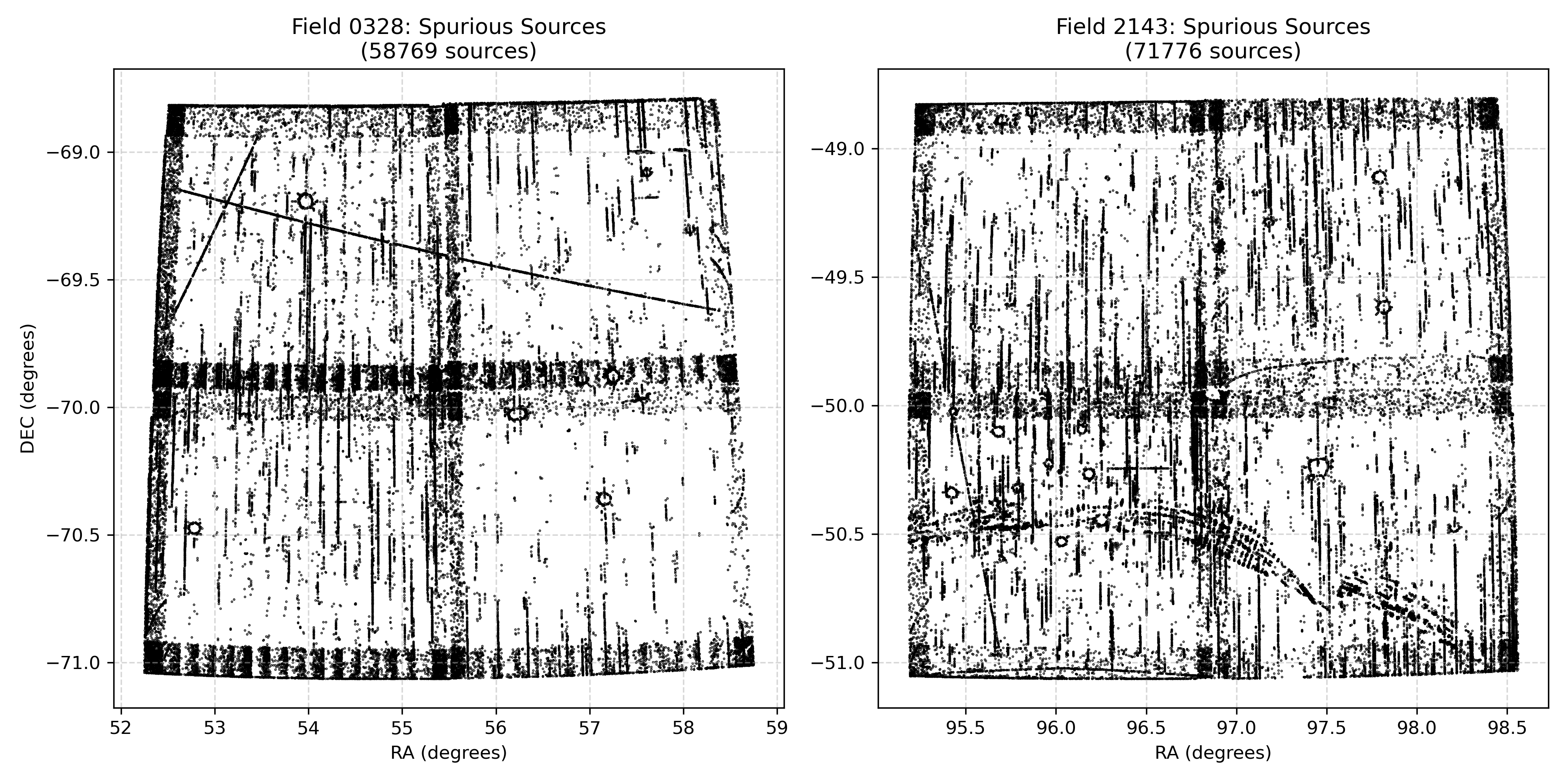}
    \caption{Examples of the spatial distribution of spurious sources in two representative KS4 tiles, 0328 (left) and 2143 (right), highlighting several archetypal artifact patterns targeted by our cleaning pipeline. These patterns include diagonal streaks from satellite trails, vertical bleeding trails from saturated stars, circular halos around bright stars, scattered features caused by stray light from bright stars located outside the field of view, and grid-like concentrations in regions of insufficient dithering.}
    \label{fig:dbscan}
\end{figure*}

\subsection{Post Processing}
\label{sec:post processing}
Following source detection and photometry, we augment the raw catalogs with supplementary information for individual sources within each tile. The main steps involve spurious source cleaning (Section~\ref{sec:removal of spurious sources}), determination of individual exposure contributions (Section~\ref{sec:tracking exposures}), and magnitude bias corrections (Section~\ref{sec:magnitude bias correction}).

\begin{figure*}
    \centering   
    \includegraphics[width=\linewidth]{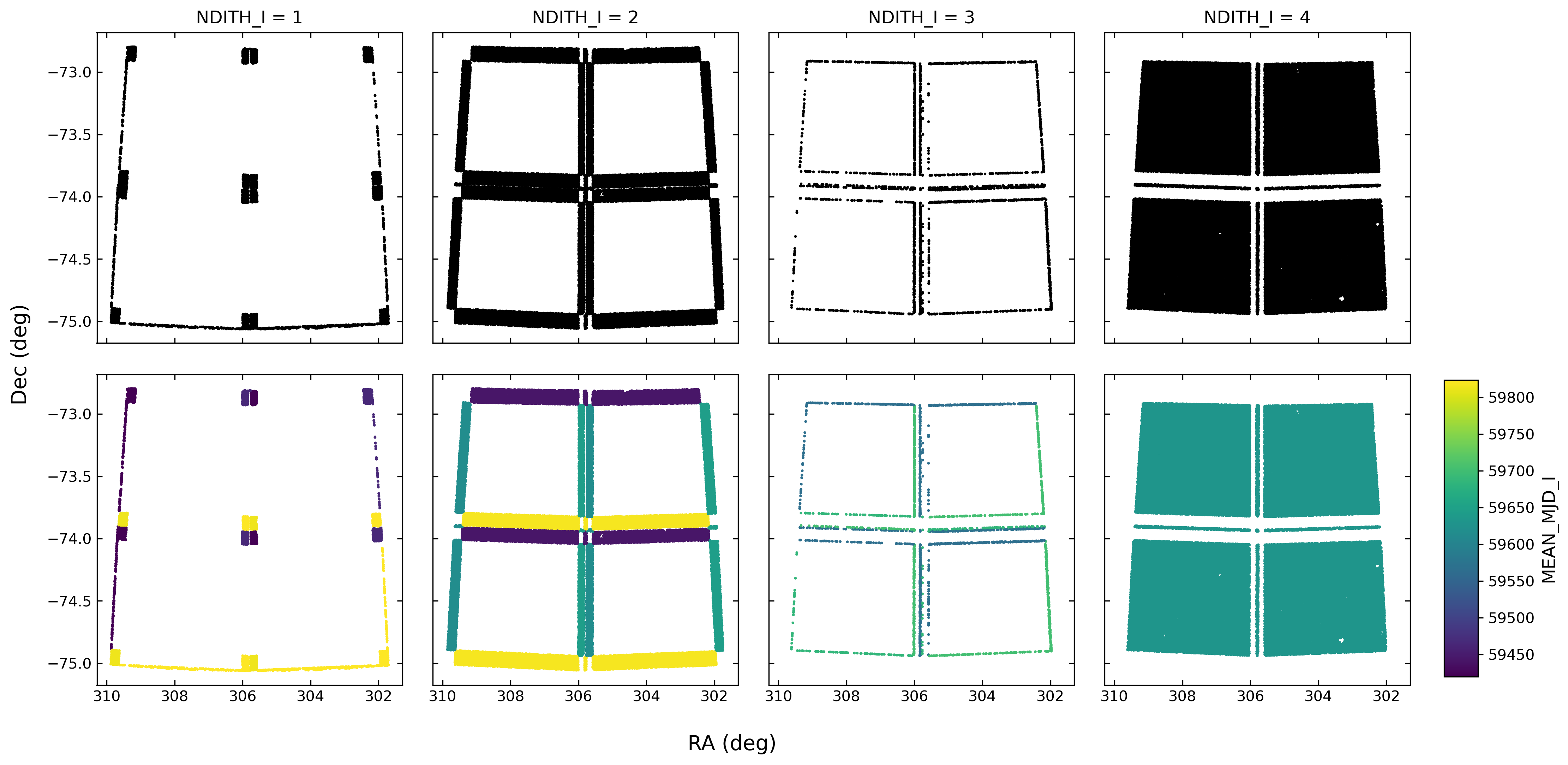}    
    \caption{Example spatial and temporal coverage patterns for a representative KS4 tile 0487. Top: The spatial distribution of detected sources (black points) within regions covered by at least \texttt{NDITH} exposures, for a tile completed with the four-point dithering strategy. White areas indicate regions where sources are not plotted because those areas received fewer than the specified \texttt{NDITH} exposures. As \texttt{NDITH} increases from 1 to 4, the panels illustrate the source distribution across progressively more uniformly covered areas. Bottom: The same sources colored by their mean observation time (\texttt{MEAN\_MJD}). The significant color variation, especially in the \texttt{NDITH} < 4 panels, highlights that some exposures were taken at widely separated epochs, in some cases separated by more than a year.}
    \label{fig:ndith}
\end{figure*}

\subsubsection{Removal of spurious sources}
\label{sec:removal of spurious sources}
The initial source extraction yields numerous spurious detections, a consequence of aggressive parameter settings intended to recover the faintest possible sources. These artifacts typically cluster in specific regions or form linear patterns associated with satellite trails and widespread stray light contamination. To eliminate these spurious detections, we implement a multi-step cleaning procedure that exploits their tendency to cluster rather than distribute randomly across the survey tile. The process begins by cross-matching sources against cataloged objects from various optical and infrared surveys to isolate a pool of unmatched candidates for further analysis.

We first define the boundary for each KS4 field using the initial source catalog positions, extending it by 3 pixels to include tile edges. We then compile a reference catalog within this boundary from multiple external surveys, including Gaia DR3 \citep{2023A&A...674A...1G}, DELVE DR2 \citep{2022ApJS..261...38D}, VHS DR6 \citep{2013Msngr.154...35M}, NSC DR2 \citep{2021AJ....161..192N}, 2MASS All-Sky Point Source Catalog \citep{2003yCat.2246....0C}, ALLWISE \citep{2019ipac.data...I1W}, SMSS DR4 \citep{2024PASA...41...61O}, and LS DR10 \citep{2019AJ....157..168D}. We cross-match KS4 sources against these known counterparts using a matching radius of 1.1 arcsec, retaining unmatched sources for spurious source analysis. We empirically determined this radius as an effective threshold to distinguish cataloged objects from potential spurious detections in our data. Next, we apply the Density-Based Spatial Clustering of Applications with Noise (DBSCAN) algorithm \citep{1996kdd...226E} to the unmatched sources to identify spatial patterns characteristic of artifacts. We configure the DBSCAN parameters (\texttt{eps}=0.03 and \texttt{min\_samples}=2) to separate genuine isolated sources from artifact clusters. Sources exhibiting clustering behavior are flagged as spurious using the \texttt{SSFLAG} parameter. Figure \ref{fig:dbscan} illustrates the typical spatial distribution of these spurious detections, which form the linear and clustered patterns. We remove all sources flagged as spurious from the final DR1, while retaining unmatched sources that show no clustering patterns (\texttt{SSFLAG=0}). Unlike the clustered patterns of the removed sources (Figure \ref{fig:dbscan}), the sources retained in the final DR1 catalog exhibit a random and isotropic spatial distribution across the tile, consistent with the expected distribution of genuine celestial objects. {\refbf The impact of this artifact removal is quantified by the False Detection Rate (FDR), defined as the ratio of spurious detections to the total number of initial detections.  Across the 979 tiles, the mean FDR per filter is $36\pm16\%$ ($B$), $31\pm14\%$ ($V$), $26\pm12\%$ ($R$), and $23\pm10\%$ ($I$).}

Although we remove spurious sources from the catalog, regions with a high density of such detections may still suffer from photometric contamination. To quantify this potential issue, we provide the \texttt{SSCOUNT\_30} parameter, which counts the number of removed spurious sources within a 30 arcsec radius of each source. Elevated values indicate regions where systematic artifacts and noise levels were likely amplified during the co-addition process. This metric allows users to assess measurement reliability and apply their own quality cuts. Users should note, however, that a high \texttt{SSCOUNT\_30} value suggests an increased potential for photometric inconsistencies rather than definitively indicating a measurement failure.

\subsubsection{Tracking Individual exposure contributions to sources}
\label{sec:tracking exposures}
Each source entry in the catalog includes metadata characterizing the spatial and temporal distribution of the contributing single-epoch exposures. The \texttt{NDITH} parameter quantifies the number of dithered exposures contributing to a source, serving as a direct proxy for survey depth uniformity. We determine this value by calculating the spatial overlap of the footprint polygons for each single-epoch image. As shown in Figure \ref{fig:ndith}, regions with \texttt{NDITH} = 4 exhibit the highest homogeneity. This parameter is therefore critical for studies demanding uniform depth, such as the KS4 quasar survey (e.g., \citealt{2024ApJS..275...46K}). Sources situated near tile edges, where polygon intersection yields unreliable results, are flagged with \texttt{EDGE = True} and assigned a conservative value of \texttt{NDITH = 1}.

We also derive temporal metadata from the observation timestamps of all contributing single-epoch exposures. The catalog provides \texttt{MIN\_MJD}, \texttt{MAX\_MJD}, and \texttt{MEAN\_MJD} for each filter. The temporal distribution is notably inhomogeneous in regions with incomplete dither coverage (\texttt{NDITH} $\leq$ 3). This heterogeneity stems from the scheduling of re-observations triggered by quality control failures. While baseline observations for a given tile were typically acquired within a single night, replacement exposures were often obtained up to a year later. This combination of epochs results in significantly extended time baselines for affected sources. Although this release provides summarized temporal statistics, it excludes individual observation timestamps for every single-epoch exposure. Therefore, the current dataset generally lacks the dense temporal sampling required for detailed time-domain variability studies.

\subsubsection{Magnitude bias correction}
\label{sec:magnitude bias correction}
We characterize systematic biases in our photometry by comparing the measurements with Gaia XP synthetic photometry \citep{2023A&A...674A...1G}. We observe magnitude-dependent offsets in \texttt{MAG\_AUTO}: bright sources ($mag \leq 15$) exhibit positive residuals, while faint sources show negative residuals relative to Gaia XP, with typical deviations of 1--3\% varying by filter and field. These offsets are negligible in fixed aperture measurements (\texttt{MAG\_APER}), suggesting they originate from the adaptive aperture algorithm of SExtractor applied to the flux-scaled co-added images.

To derive empirical corrections, we first quantify this bias relative to the robust aperture photometry. {\refbf Since these trends depend on both filter and tile, a unique empirical correction function is derived independently for each filter-tile combination.} We select a sample of reliable point sources matched to Gaia XP, applying specific selection criteria: \texttt{CLASS\_STAR} $> 0.8$, \texttt{FLAGS} $< 4$, and isolation within 5 arcsec. Using this clean sample, we calculate the magnitude difference $\Delta m = \texttt{MAG\_APER10} - \texttt{MAG\_AUTO}$ as a function of apparent magnitude (see Figure \ref{fig:magbias}). We adopt adaptive magnitude bins ranging from 0.125 to 0.25 mag to ensure adequate sampling across the full dynamic range. To construct the correction function, we compute the median value for each bin, thereby rejecting outliers. The final function is generated by linearly interpolating between these discrete values. {\refbf A typical example of such a correction function is presented in the left panel of Figure \ref{fig:magbias}.} We apply statistical weighting based on source density during this interpolation to suppress noise while preserving the underlying magnitude-dependent trend.

We apply these stellar-calibrated corrections to two source categories: all Gaia-matched objects used in the calibration and high-confidence point sources (\texttt{CLASS\_STAR} $> 0.9$) lacking Gaia counterparts. Extended sources and low-confidence point sources remain uncorrected, primarily due to the lack of reliable reference magnitudes for galaxies to validate such corrections.

\begin{figure*}
    \centering
    \includegraphics[width=\linewidth]{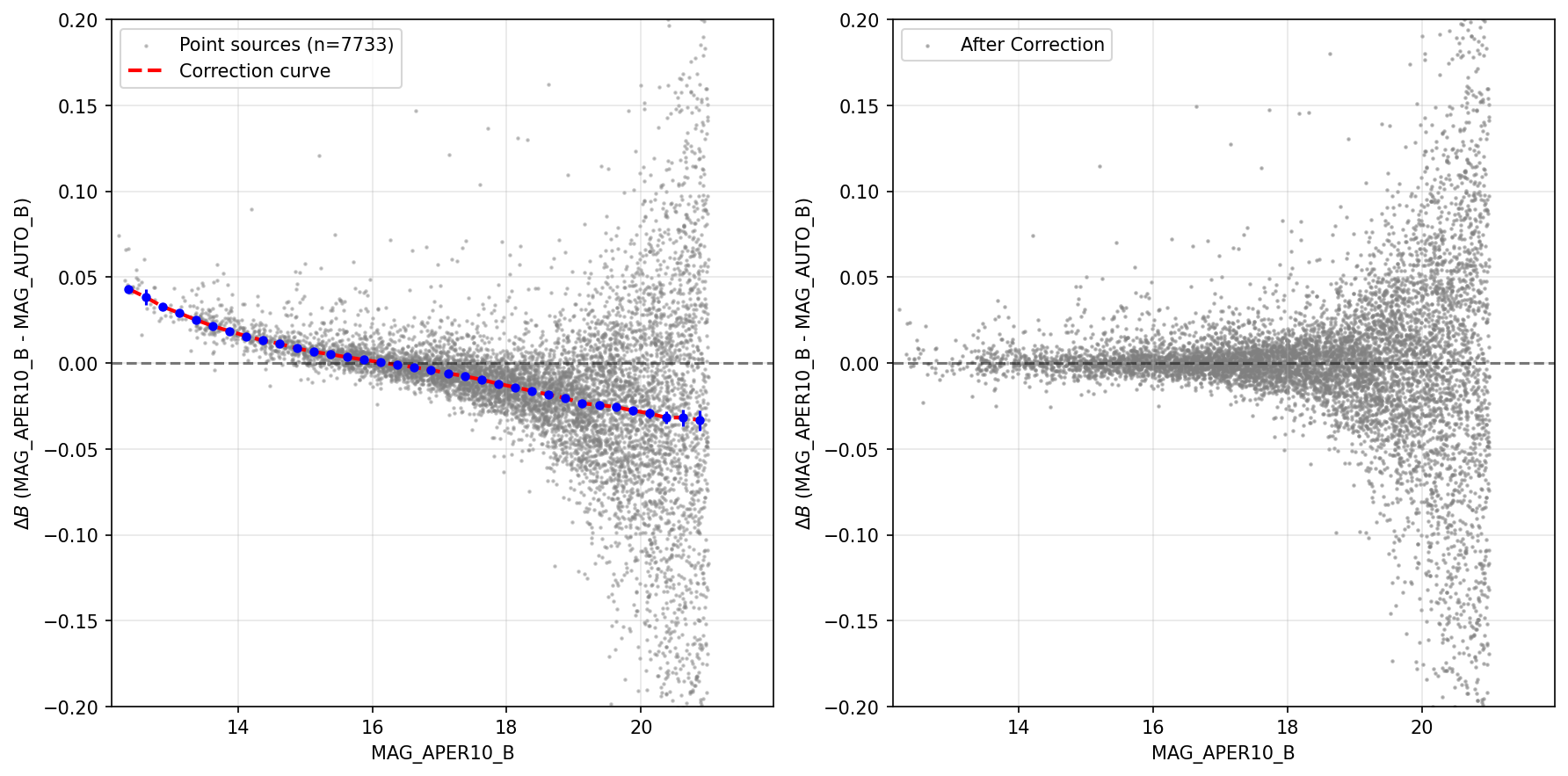}
    \caption{An example of the empirical correction for the magnitude-dependent bias observed in \texttt{MAG\_AUTO}. The plots show the magnitude difference ($\Delta B$ = \texttt{MAG\_APER10\_B} - \texttt{MAG\_AUTO\_B}) for reliable stellar sources in the $B$-band of KS4 tile 2747, cross-matched with GAIA XP. Left: The distribution before correction, showing a clear systematic trend with magnitude. The red dashed line is the derived correction function. Right: The distribution after the correction is applied, showing the successful removal of the systematic bias.}
    \label{fig:magbias}
\end{figure*}


\section{Data Products}
\label{sec:data products}
KS4 DR1 provides science-ready data products designed to enable a wide range of astronomical research. The release includes processed imaging data (deep co-added images, single-epoch exposures, and masks), Hierarchical Progressive Survey (\citealt{2015A&A...578A.114F}) datasets, photometric source catalogs, and ancillary zero-point correction maps. The following subsections provide detailed descriptions of each product.

\subsection{Images}
\label{sec:images}
The primary data products are the deep, co-added images, comprising 6.8 TB of data covering 979 distinct fields in the $BVRI$ bands. Each co-added image is generated from multiple 120-second exposures (see Table~\ref{tab:exposure_distribution}), processed to achieve the final survey depth, and provided as large-format ($22,000 \times 22,000$ pixels) files. These are accompanied by corresponding mask images (1.7 TB) containing essential quality-control information for source flagging. We also release the full set of single-epoch images, which constitutes the bulk of the data volume at 27.5 TB (22.0 TB of science images and 5.5 TB of associated masks). Unlike the co-added images, each single-epoch observation is distributed as a set of four individual CCD frames ($9216 \times 9232$ pixels each). These individual exposures retain the original temporal information, enabling investigations of variable and moving objects. However, the temporal sampling is constrained by the survey strategy. A typical observation sequence comprises four or more exposures clustered within a few minutes, whereas some visits are separated by extended intervals of approximately one year. Thus, the dataset is particularly sensitive to either very short-term or long-term variability.

To utilize the interactive visualization and rapid exploration of this multi-terabyte dataset, we provide data products in the HiPS format. The HiPS maps were generated using Hipsgen\footnote{Developed and maintained by the Strasbourg astronomical Data Center; \url{https://aladin.cds.unistra.fr/hips/}} to a maximum HEALPix order of 10. This configuration yields a native pixel resolution of 0.403 arcsec (402.6 mas), with data encoded as 8-bit PNG tiles. These HiPS products are available for individual $BVRI$ filters as well as a three-color ($BVI$) composite. We also provide a MOC map defining the precise survey footprint, which covers 4,044 deg$^{2}$ (9.8\%) of the celestial sphere. The color mapping for the composite image was optimized using an asinh stretch to enhance dynamic range, resulting in a visual appearance comparable to that of the Legacy Survey DR10 {\refbf \citep{2019AJ....157..168D}}.


\subsection{Photometric Source Catalogs}
\label{sec:catalogs}
We present two distinct photometric source catalogs, each tailored to specific scientific applications: a reference-band catalog based on $I$-band detections (\texttt{idual\_master}) and a band-merged catalog constructed from independent single-band detections (\texttt{single\_master}).

\subsubsection{The $I$-band Reference Catalog}
\label{sec:iband reference catalog}
This catalog is specifically designed to support the key science drivers of KS4, including searches for high-redshift quasars and galaxy clusters, which rely on robust color information for faint objects. Accordingly, we adopt the deep $I$-band co-added image as the reference for detection and perform forced photometry on the $B$, $V$, and $R$ images at these fixed positions. This dual-image mode strategy eliminates the need for inter-band cross-matching, ensuring homogeneous photometry while providing meaningful flux estimates or upper limits for sources otherwise undetected in bluer bands. As a practical validation, an earlier version of this reference catalog from the EDR was successfully utilized for the selection of KS4 quasars (see \citealt{2024ApJS..275...46K}).

The final dataset, \texttt{idual\_master}, is constructed by aggregating multiple measurements for each unique source. This aggregation is particularly critical for handling objects detected in the overlapping regions between adjacent tiles. Our strategy adopts the robust procedures developed for the SkyMapper DR4 catalog \citep{2024PASA...41...61O}, an approach well-suited to surveys with a limited number of epochs. We define unique sources by linking individual detections situated within a 0.5 arcsec radius. This radius accommodates the degraded astrometric precision often observed near tile edges, which affects the majority of sources in overlap regions. Each source group is assigned a unique \texttt{COADD\_OBJECT\_ID}, structured following the Gaia identifier scheme \citep{2023A&A...674A...1G}. This 64-bit integer encodes the approximate source position by embedding a level 12 nested HEALPix index in the most significant bits, a format we adopt to enable highly efficient spatial queries. For convenience, the level 12 index is also provided in a separate \texttt{HEALPIX\_INDEX} column. 

Once unique sources are identified, the final table is generated by aggregating the multiple available measurements for each source. This process incorporates only individual detections satisfying the following criteria: \texttt{SNR} $> 2$ and \texttt{FLAGS} $< 4$. These inclusive selection criteria mirror the strategy adopted by DELVE \citep{2021ApJS..256....2D}. The \texttt{SNR} $> 2$ threshold retains even low-significance detections, whereas the \texttt{FLAGS} $< 4$ cut excludes saturated detections (\texttt{FLAGS} $= 4$) while retaining sources with minor warnings. This approach provides a more complete catalog, allowing users to apply stricter \texttt{SNR} or \texttt{FLAGS} cuts as needed for high-purity samples. From this clean set of measurements, the final parameters for each source are derived as follows:

\begin{itemize}
    \item Photometry: The final catalog magnitude, for both automatic (\texttt{MAG\_AUTO}) and aperture measurements, is calculated as the weighted median of selected detections, using weights of $1/\sigma^{2}$ where $\sigma$ is the photometric error. The associated uncertainty is derived via standard error propagation.
  
    \item Morphology: Core morphological parameters (e.g., \texttt{CLASS\_STAR}, \texttt{ELLIPTICITY}, \texttt{A\_IMAGE}, \texttt{B\_IMAGE}, and \texttt{FLUX\_RADIUS}) are adopted from the single detection exhibiting the smallest FWHM (i.e., the best seeing) within each source group, which typically comprises 2–3 detections (up to 5). In contrast, the cataloged \texttt{FWHM\_IMAGE} represents the mean FWHM of all reliable detections for that source.
  
    \item Other Parameters: Observational timestamps (\texttt{MJD}) are summarized by their minimum, maximum, and mean values. Quality flags are combined conservatively to ensure all potential issues are recorded: \texttt{FLAGS} and \texttt{IMAFLAGS\_ISO} are merged using a bitwise \texttt{OR}, while the \texttt{EDGE} flag is merged using a logical \texttt{OR}. For parameters where the maximum value represents the most critical information, such as the number of dithers (\texttt{NDITH}) and spurious source count (\texttt{SSCOUNT\_30}), the maximum is selected across all detections. Finally, the total number of individual detections comprising each group is recorded in the \texttt{N\_DET} column.
    
\end{itemize}



This catalog contains 228,119,100 unique sources. Of these, 201,912,335 (88.5\%) correspond to single-tile detections, while the remaining 26,206,765 (11.5\%) represent sources identified in the overlap regions between adjacent tiles. The catalog is further supplemented with value-added data, including estimates of line-of-sight Galactic reddening, $E(B-V)$, derived from the dust maps of \citet{1998ApJ...500..525S} via the public \texttt{dustmap} tool \citep{2018JOSS....3..695G}.

\begin{figure*} 
    \centering
     \includegraphics[width=\linewidth]{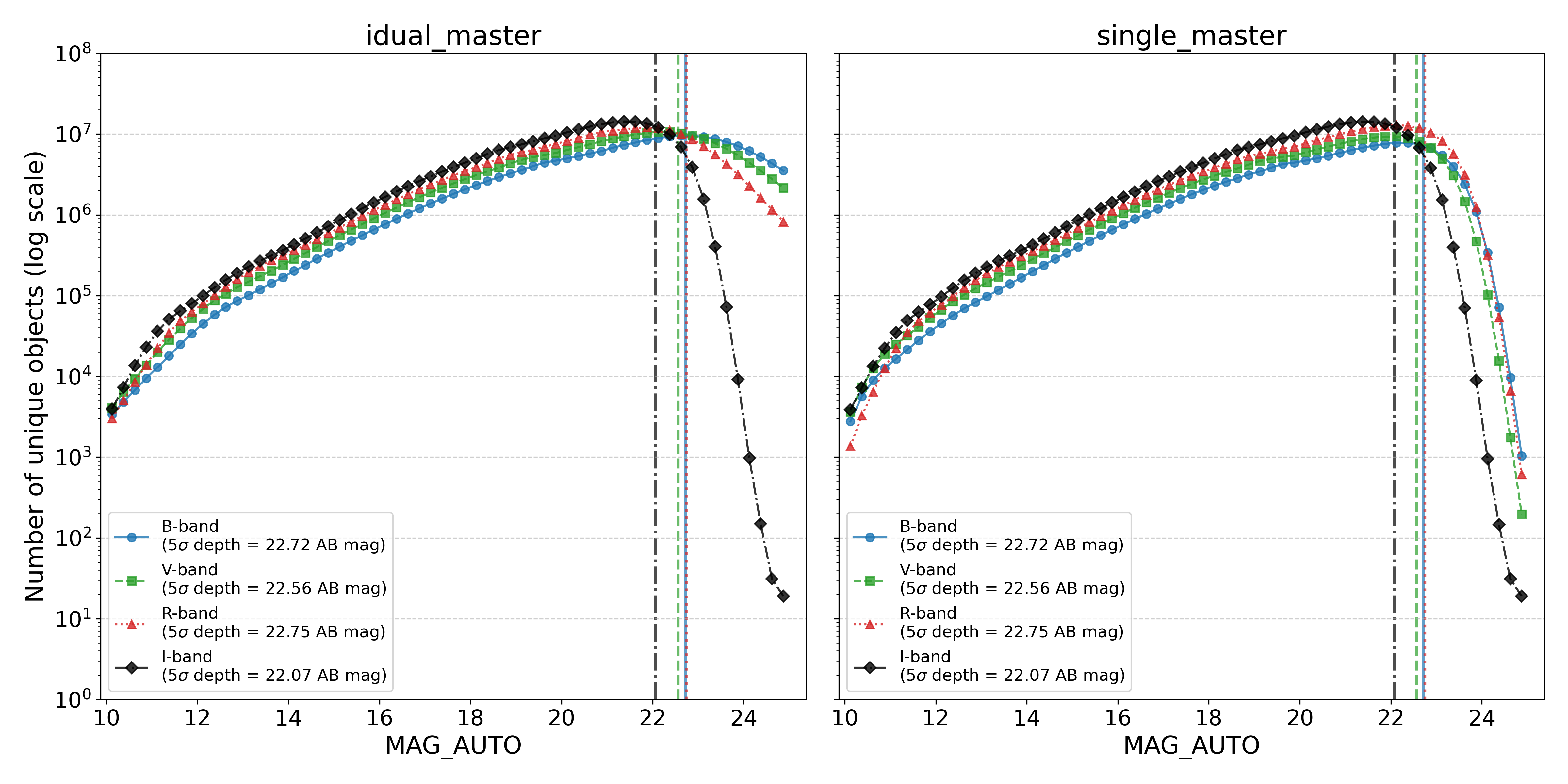}
     \caption{Magnitude distributions of sources in the \texttt{idual\_master} (left) and \texttt{single\_master} (right) catalogs. The histograms display the number of unique objects as a function of \texttt{MAG\_AUTO} for the $B$, $V$, $R$, and $I$ bands, representing all cataloged sources without an SNR cut. The vertical lines mark the median $5\sigma$ limiting magnitudes derived in \citep{2026arXiv2603.17442J}, corresponding to the threshold for high-confidence detections.}     
    \label{fig:magnitude distribution}
\end{figure*}

\subsubsection{The Band-Merged Catalog}
\label{sec:band-merged catalog}
We construct the band-merged \texttt{single\_master} catalog to incorporate all objects reliably detected in at least one of the $BVRI$ bands. The process begins with independent source detection on each of the four filter images using SExtractor \citep{1996A&AS..117..393B}. We then merge the resulting single-band source lists into a master catalog via a hierarchical cross-matching procedure, applying a spatial matching radius of 0.8 arcsec. We empirically determined this radius based on tests using KMTNet observations of the COSMOS field \citep{2018JKAS...51...89K}; the abundant multi-wavelength data available in this field provided an ideal benchmark for validating the matching accuracy of extended sources. The matching process proceeds in a specific sequence: the $R$- and $I$-band catalogs are first combined, followed by the $V$-band list, and finally merged with the $B$-band catalog. {\refbf The median separations are $0.118\pm0.157$ arcsec ($R$--$I$), $0.104\pm0.148$ arcsec ($V$--$RI$), and $0.100\pm0.146$ arcsec ($B$--$VRI$), confirming sub-pixel alignment.} For sources detected in multiple bands, the final celestial coordinates are calculated as the unweighted mean of the individual detections, with their associated uncertainties combined in quadrature. Unlike the reference catalog, which relies on an $I$-band selection, the \texttt{single\_master} catalog comprises all sources satisfying an \texttt{SNR} $> 2$ in any of the $BVRI$ bands. For bands where a source is not detected, the corresponding photometric values are assigned a null value. We distinguish between non-detections (observed but not detected) and non-observations; the latter is explicitly indicated by \texttt{NDITH} = 0. For non-detections, we provide the corresponding $5\sigma$ image depth via the public repository\footnote{\url{https://github.com/jmk5040/KMTNet_ToO/blob/main/pipe/KS4DR1_GetDepth.py}} to allow for upper limit estimation. The subsequent aggregation procedures follow those described for the \texttt{idual\_master} catalog (see Section~\ref{sec:iband reference catalog}).

The band-merged strategy of the \texttt{single\_master} catalog substantially improves completeness by recovering sources missed by the $I$-band-limited selection of the \texttt{idual\_master} catalog. This advantage is particularly evident for blue objects, where the independent detection strategy effectively identifies sources that are faint in the $I$-band. For instance, the catalog includes 3,732,594 sources detected in the $B$, $V$, and $R$ bands but undetected in the $I$-band, along with an additional 2,109,238 sources found exclusively in the $B$ and $V$ bands. Given its enhanced completeness, the \texttt{single\_master} catalog serves as the primary resource for the broader astronomical community. The final catalog includes a total of 279,909,360 unique entries. Of these, 248,360,987 (88.7\%) correspond to single-tile detections, while the remaining 31,548,373 (11.3\%) represent overlapping sources identified in adjacent tile regions. Regarding high-confidence detections (SNR $> 5$), the catalog contains 116,393,930 sources in the $B$-band, 140,896,979 in $V$, 195,153,754 in $R$, and 202,256,982 in $I$.

Figure \ref{fig:magnitude distribution} highlights the distinct magnitude distributions of the two catalogs. The \texttt{single\_master} catalog exhibits a characteristic turnover near the detection limit, whereas the forced photometry approach of the \texttt{idual\_master} catalog yields $BVR$ distributions extending to fainter magnitudes. {\refbf Conversely, the nearly identical $I$-band distributions confirm that the sequential merging and cleaning procedures preserved the depth and completeness of the reference band.}



\begin{figure*}[!h]
    \centering
    \includegraphics[width=\linewidth]{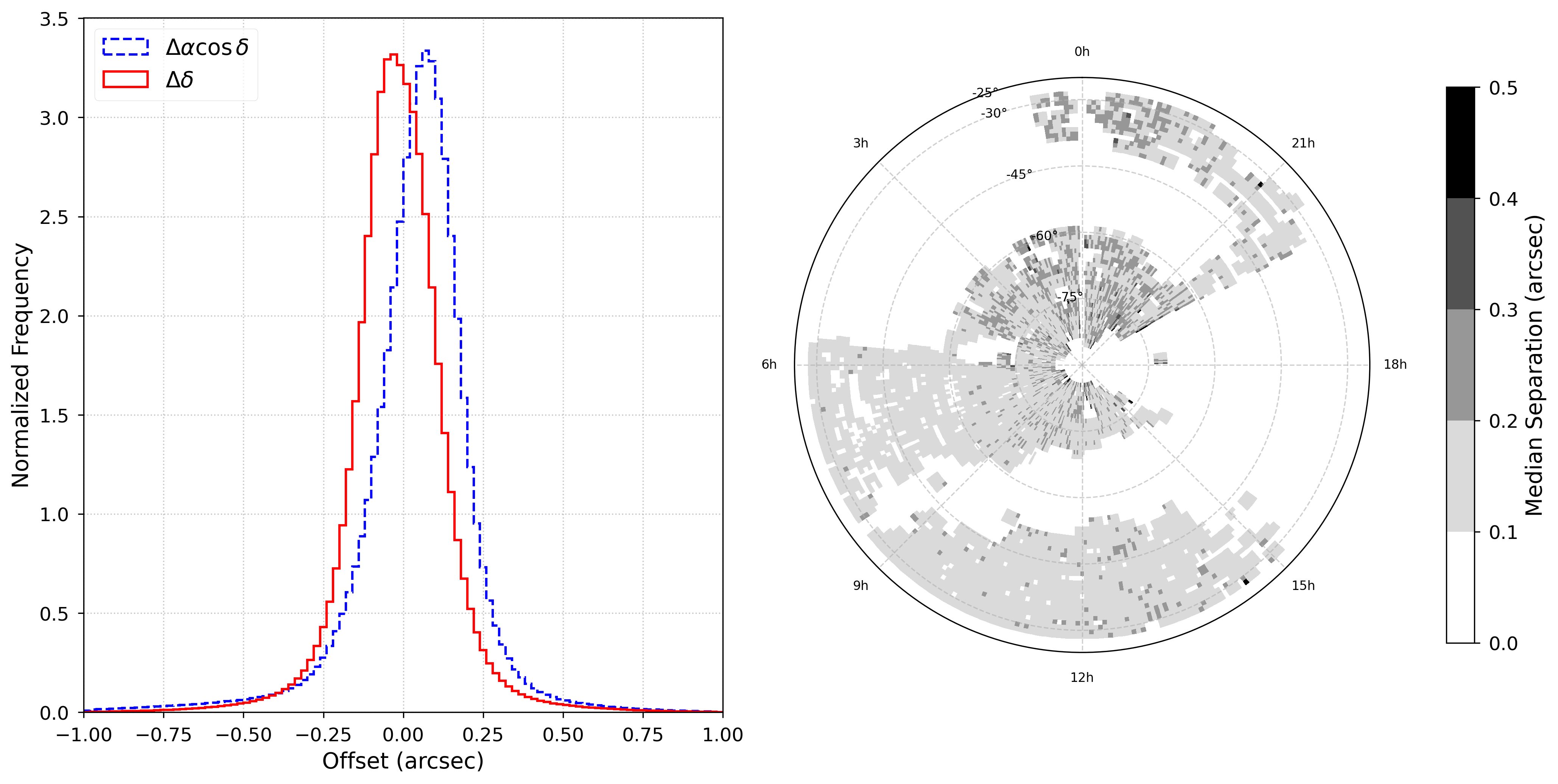}
    \caption{Astrometric performance of the KS4 DR1 catalog against Gaia DR3. 
    \textit{Left}: Normalized histograms of the astrometric offsets, showing the distributions for RA ($\Delta\alpha \cos\delta$, dashed line) and Dec ($\Delta\delta$, solid line). The mean offsets for the dRA and dDec components are $+0.054 \pm 0.129$ arcsec and $-0.015 \pm 0.120$ arcsec, respectively.
    \textit{Right}: Polar skymap showing the median separation in 1-degree bins across the DR1 footprint for sources with separations < 0.5 arcsec.}
\label{fig:astrometry_fig1}
\end{figure*}

\subsection{Zero-Point Correction Maps}
\label{sec:zp correction map}
We release ancillary two-dimensional ZP correction maps as part of the DR1 data products. {\refbf These maps are provided as 2D grids with a spatial bin size of $400 \times 400$ pixels. They encode the additive magnitude corrections required within each bin to compensate for local ZP spatial variations. While the primary DR1 catalogs already incorporate these corrections for standard apertures, these maps are essential for users performing precise custom photometry. The local ZP correction is obtained by identifying the $400 \times 400$ pixel bin corresponding to the center (X, Y) of a source and applying the constant magnitude offset assigned to that specific bin. A comprehensive description of the map generation and validation procedures is provided in Section 3.6.1 and Figure 7 of \citet{2026arXiv2603.17442J}.}


For each tile, we provide a ZP map (\texttt{zp\_map}) and a corresponding uncertainty map (\texttt{zperr\_map}). These are generated for each of the primary photometric apertures (3, 5, and 10 arcsec, and \texttt{MAG\_AUTO}) adopted in our catalogs. Users can retrieve the precise local ZP correction by sampling these maps at specific pixel coordinates (X, Y). Applying these corrections substantially improves photometric uniformity; for a typical field, the photometric RMSE relative to Gaia decreases from $\sim0.028$ mag to $\sim0.019$ mag. Furthermore, the \texttt{zperr\_map} enables users to impose quality cuts by excluding sources located in regions where the uncertainty exceeds a specified threshold.

\begin{figure}[!t]
    \centering
    \includegraphics[width=\linewidth]{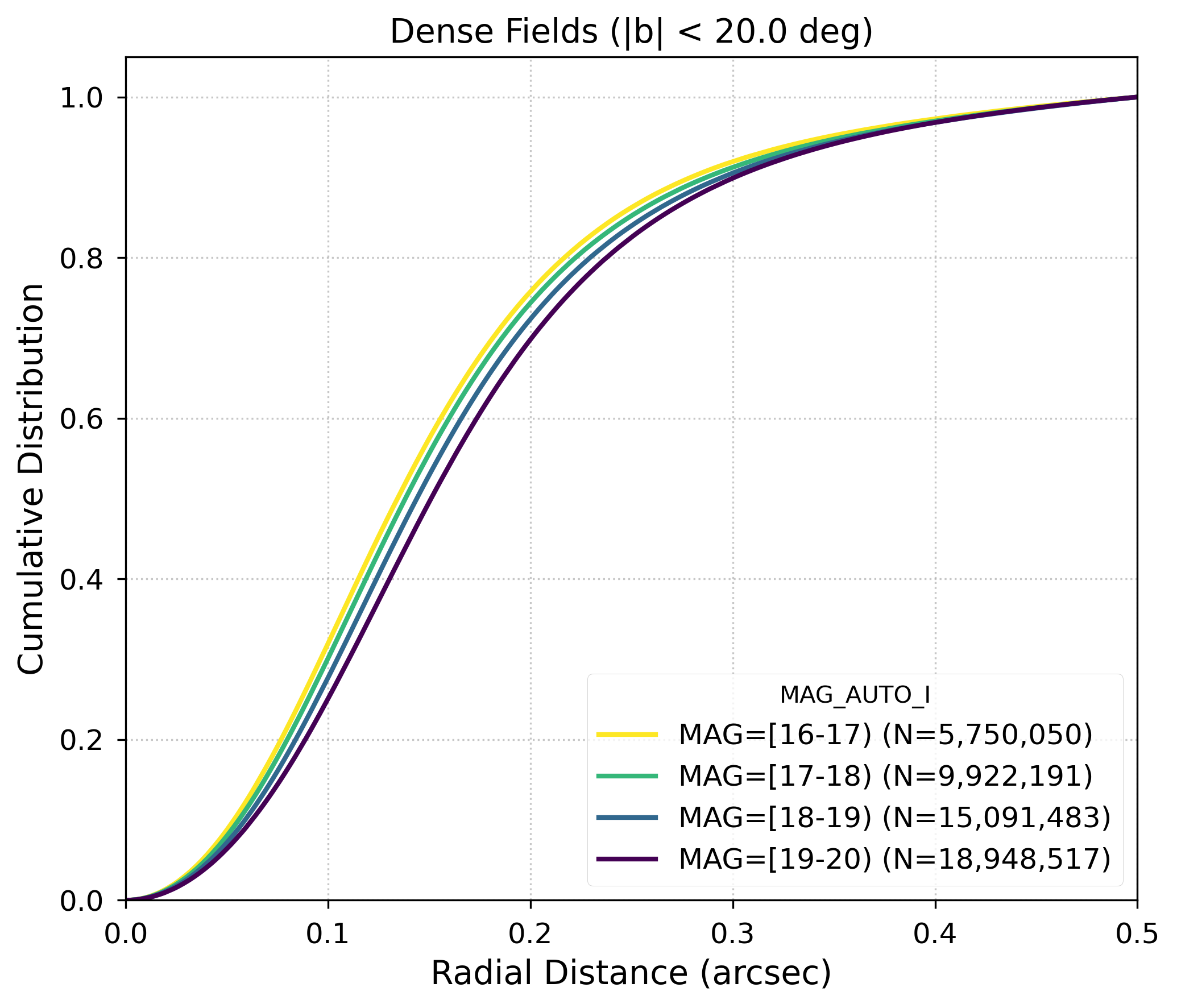}
    \includegraphics[width=\linewidth]{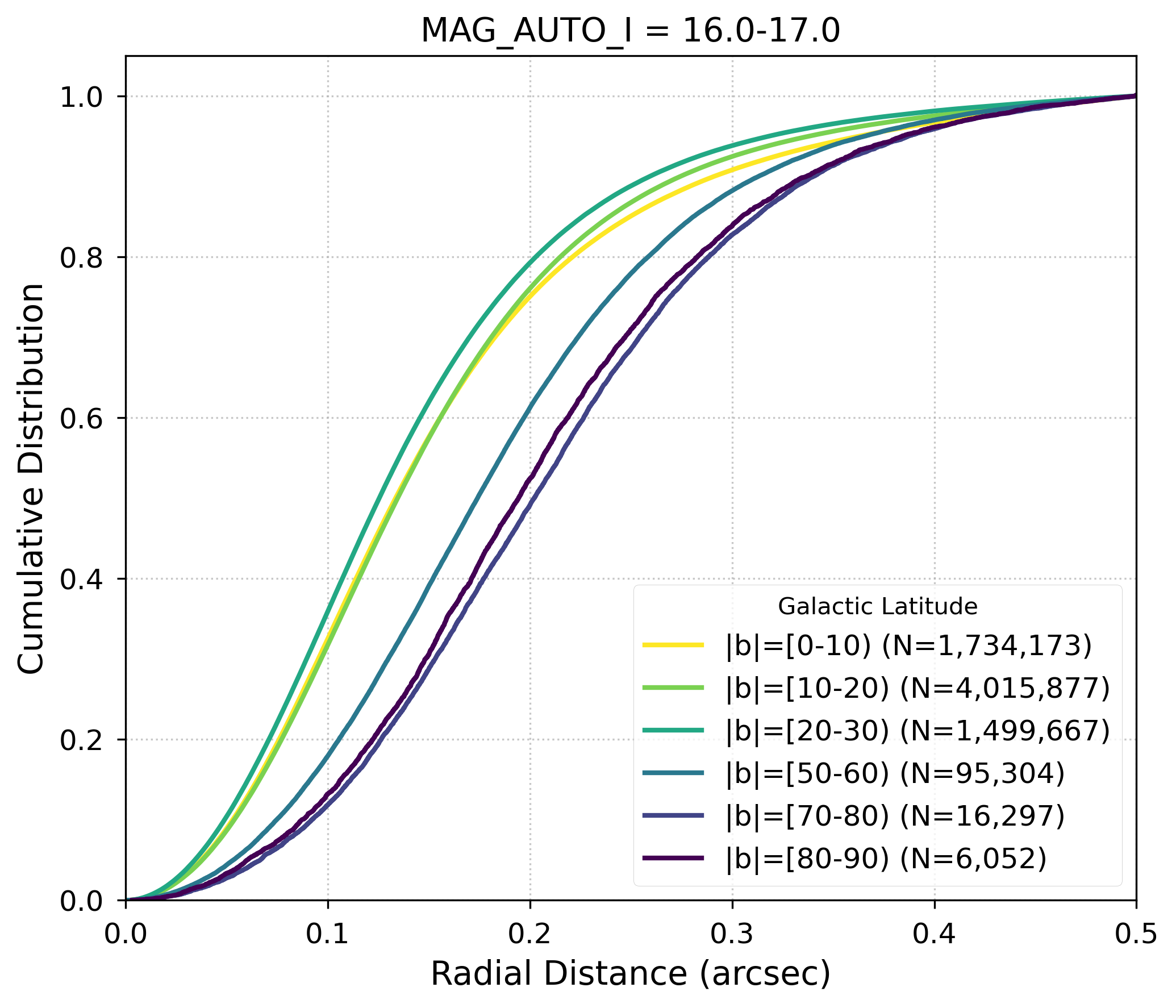}
    \caption{Cumulative distribution functions of  the radial astrometric offset between the KS4 \texttt{idual\_master} catalog and Gaia DR3. 
    \textit{Top}: The astrometric precision in dense fields ($|b| < 20$ deg), shown for different $I$-band magnitude bins. 
    \textit{Bottom}: The astrometric precision as a function of absolute Galactic latitude ($|b|$) for a sample of bright stars with $16 < \texttt{MAG\_AUTO\_I} < 17$.}
    \label{fig:astrometry_cdf}
\end{figure}

\section{Data Validation and Quality Assessment}
\label{sec:data validation and quality assessment}

\subsection{Astrometric Accuracy and Precision}
\label{sec:astrometric performance}
We evaluate the astrometric performance of DR1 by cross-matching the \texttt{idual\_master} catalog against Gaia DR3. We selected the \texttt{idual\_master} catalog for this validation because its source positions provide a single, unambiguous reference, unlike the multi-band averaged coordinates found in the \texttt{single\_master} catalog. Despite the temporal baseline between our observations and the Gaia DR3 reference epoch (J2016.0), we proceeded with a direct positional cross-match using a 1.0 arcsecond radius. We did not apply proper motion corrections; As a result, the matched sample is inherently biased toward sources with low proper motions. In total, the cross-match yields 115,156,855 associations with Gaia DR3. The left panel of Figure \ref{fig:astrometry_fig1} presents the normalized distributions of astrometric residuals in RA ($\Delta\alpha \cos\delta$) and Dec ($\Delta\delta$). These distributions confirm the lack of significant systematic bias, exhibiting mean offsets of $+0.054\pm0.129$ arcsec and $-0.015\pm0.120$ arcsec, respectively. Separately, a calculation of the total radial separation yields a modal offset of 0.125 arcsec (corresponding to $\sim31\%$ of the KMTNet pixel scale), indicating tight alignment. The right panel illustrates the spatial distribution of median separations binned at 1-degree resolution, demonstrating high uniformity across the DR1 footprint. These results confirm that the KS4 catalog achieves high astrometric accuracy relative to the Gaia DR3 reference frame, particularly for stationary sources.

We investigate the dependence of astrometric precision on source brightness and stellar density in Figure \ref{fig:astrometry_cdf}. The top panel displays the performance in dense fields ($|b|<20^{\circ}$) using cumulative distribution functions (CDFs) of radial offsets, subdivided by magnitude bins. As expected, astrometric precision strongly correlates with source brightness; the CDF for the brightest sources exhibits the steepest ascent, while the curves for fainter sources systematically shift toward larger separations. This trend aligns with the expected degradation of centroiding accuracy as the SNR decreases. In the bottom panel, the dependence on Galactic latitude is examined for a magnitude-limited sample of bright stars ($16 < \texttt{MAG\_AUTO\_I} < 17$). A subtle yet systematic trend is observed where precision is slightly degraded in sparse, high-latitude fields compared to crowded regions near the Galactic plane. This behavior is attributed to the reduced density of Gaia DR3 reference stars at high latitudes, which results in less robust constraints on the astrometric solution.

\begin{figure*}[!h]
    \centering
    \includegraphics[width=\linewidth]{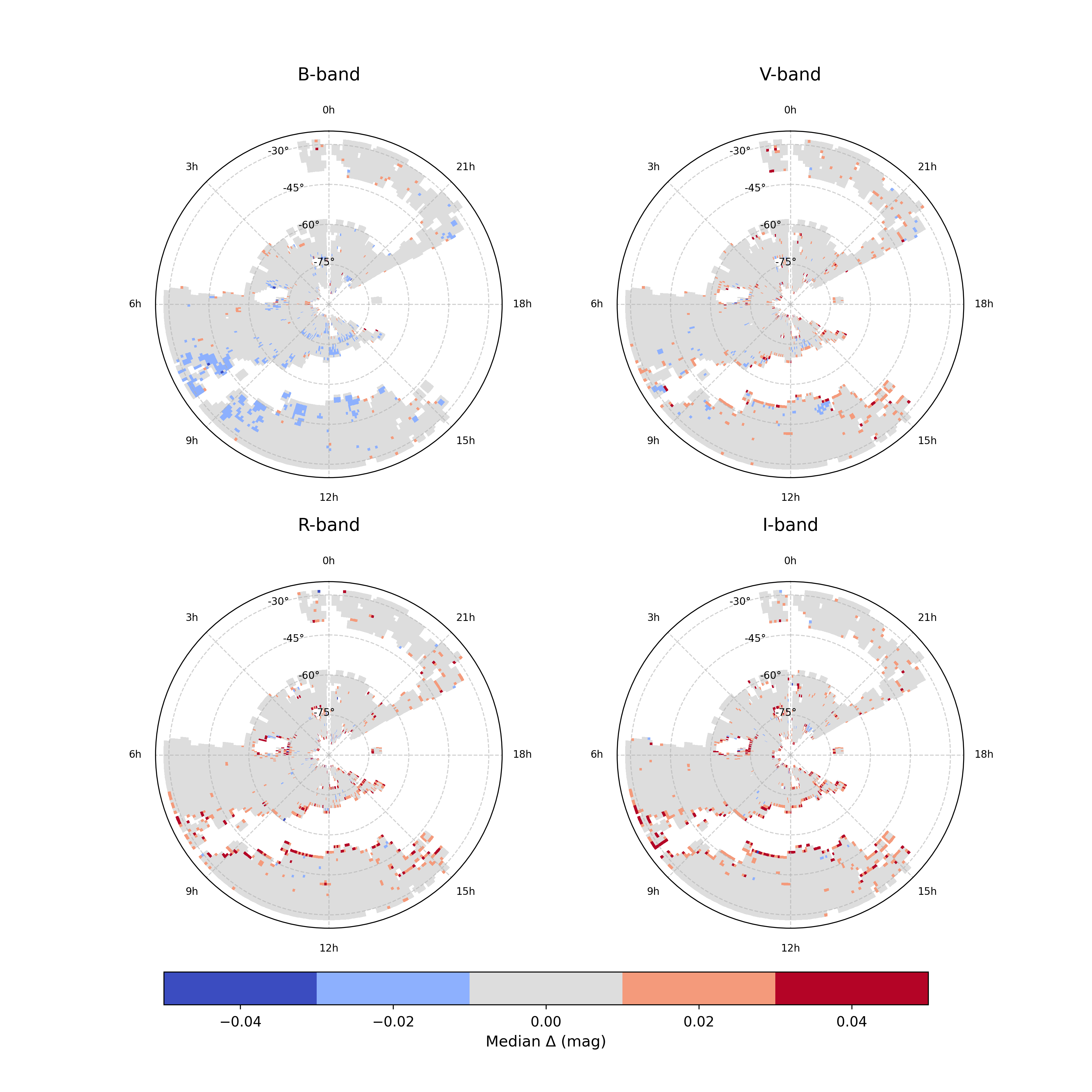}
    \caption{Skymaps showing the median photometric offset ($\Delta m$ = $m_{GaiaXP} - m_{KS4}$) between the KS4 DR1 \texttt{MAG\_AUTO} photometry and Gaia XP synthetic photometry for a large sample of point sources (\texttt{CLASS\_STAR} > 0.9). For the B-band comparison, the Gaia XP magnitudes were color-term corrected. The offsets for each 1-degree bin are calculated from a 3-$\sigma$ clipped distribution. No correction for Galactic extinction has been applied to the magnitudes used in this comparison. The four panels show the results for the $B$ (top left), $V$ (top right), $R$ (bottom left), and $I$ (bottom right) bands in a polar projection. Red indicates regions where KS4 magnitudes are systematically brighter than Gaia XP, while blue indicates regions where they are fainter.}
    \label{fig:photometry_gaiaxp}
\end{figure*}

\begin{figure*}[!h]
    \centering
    \includegraphics[width=\linewidth]{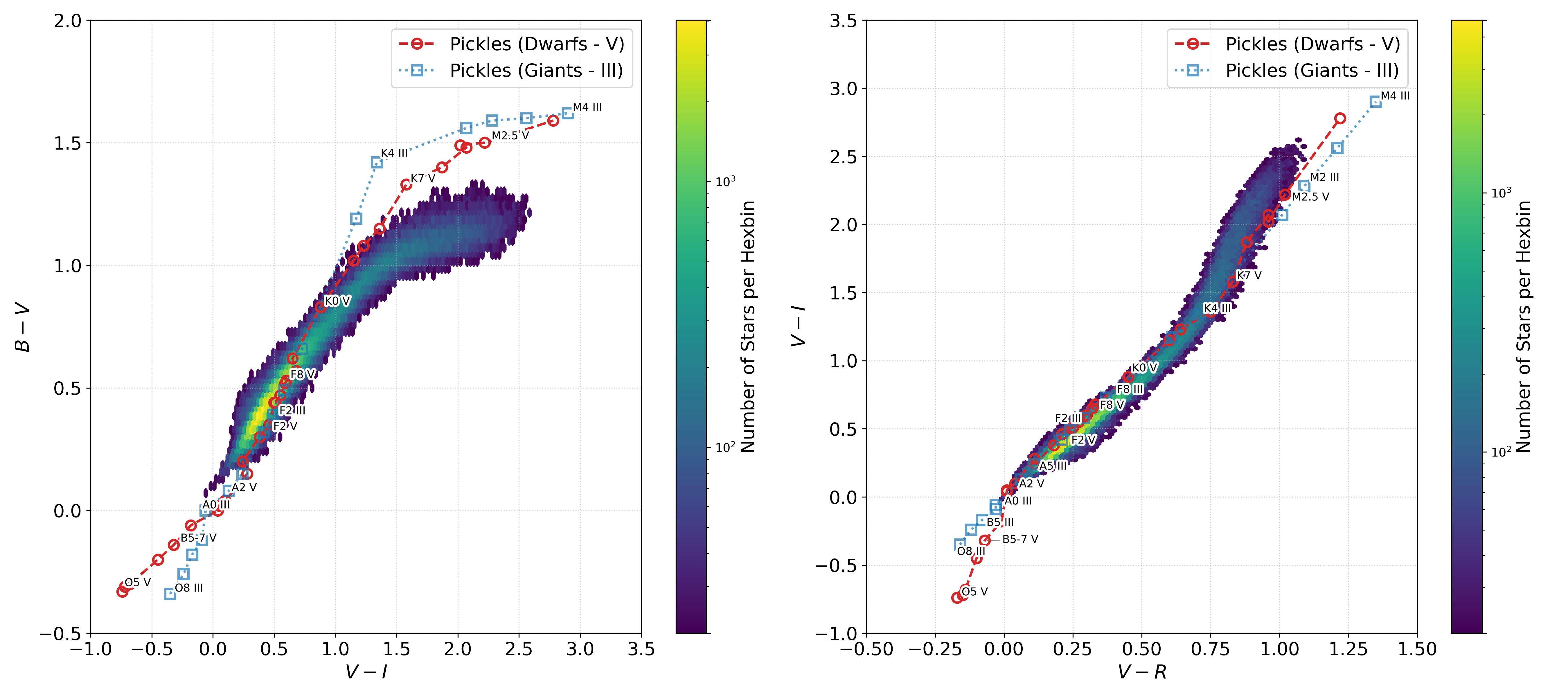}
    \caption{Color-color diagrams showing the stellar loci of the KS4 photometric system. The locus is shown as a 2D density distribution of high-quality point sources from the \texttt{single\_master} catalog, selected from regions of low Galactic extinction ($E(B-V)$ < 0.2); no correction for extinction has been applied. Overplotted for comparison are stellar loci representing the traditional Johnson-Cousins system, derived from the \citet{1998PASP..110..863P} library for dwarf (V) and giant (III) stars.   \textit{Left}: The $(B-V)$ vs. $(V-I)$ diagram. \textit{Right}: The $(V-I)$ vs. $(V-R)$ diagram.}
    \label{fig:stellar locus}
\end{figure*}

\subsection{Photometric Performance}
\label{sec:photometric performance}
We evaluate the photometric uniformity of point sources across the DR1 footprint by comparing our \texttt{MAG\_AUTO} photometry against Gaia XP synthetic photometry \citep{2023A&A...674A...1G}. To create a comprehensive sample for this assessment, we compile millions of cross-matched sources spanning the full magnitude range available in each filter. This extensive dataset provides a more robust validation of the final data products compared to the analysis presented in \citet{2026arXiv2603.17442J}, which relied on a restricted, signal-to-noise-limited sample ($14 < mag < 17$) for generating local zero-point deviation maps. The validation sample contains high-quality point sources selected via strict criteria (\texttt{FLAGS}=0, \texttt{EDGE}=FALSE, \texttt{IMAFLAGS\_ISO}=0, \texttt{CLASS\_STAR} $> 0.9$) and cross-matched to Gaia XP within a 0.5 arcsec radius. For each band ($B, V, R, I$), we compute the magnitude difference, $\Delta m = m_{\rm GaiaXP} - m_{\rm KS4}$. We emphasize that this comparison uses observed magnitudes without correction for Galactic extinction. For the $B$-band, we apply a color-term correction to the Gaia XP magnitude: $B_{\rm corr} = B_{\rm GaiaXP} - 0.3 \times (B_{\rm GaiaXP} - V_{\rm GaiaXP})$. We empirically determined this coefficient to minimize residual scatter against KS4 magnitudes (see \citealt{2026arXiv2603.17442J} for details). The resulting $B$-band residual is defined as $\Delta B = B_{\rm corr} - B_{\rm KS4}$. To reject outliers, we apply a 3$\sigma$ clipping algorithm to the $\Delta m$ distribution before calculating the median offset within 1-degree spatial bins. {\refbf Figure \ref{fig:photometry_gaiaxp} presents the sky maps of median photometric offsets for all four bands. These maps demonstrate high photometric homogeneity, with the fraction of the footprint maintaining median offsets relative to Gaia XP within $\pm0.03$ mag being 99.8\%, 99.1\%, 98.0\%, and 97.5\% for the $B, V, R,$ and $I$ bands, respectively. Notably, over 92\% of the regions in all bands achieve even tighter uniformity with offsets $\le \pm0.01$ mag, as indicated by the predominantly gray shading in the figure. The $R$ and $I$ bands exhibit the highest stability, yielding typical median absolute offsets of $\sim$0.001 mag.} Minor systematic deviations are observed primarily near tile boundaries, particularly in regions adjacent to low Galactic latitudes. We attribute these residuals to differential extinction effects between the KS4 and Gaia filter systems, which are most pronounced in high-extinction areas.

To validate the internal consistency of the multi-band photometry, we examine stellar locus diagrams. This validation is crucial given that the KMTNet filter transmission curves deviate from the standard Johnson-Cousins system. While a global color term correction was applied during calibration against Gaia XP, the stellar locus offers a more comprehensive assessment of the final color accuracy. For this analysis, we select a clean sample using strict quality cuts: \texttt{FLAGS}=0, \texttt{IMAFLAGS\_ISO}=0, \texttt{EDGE}=False, deep coverage ($\texttt{NDITH} > 3$), low Galactic reddening ($E(B-V)_{\rm SFD} < 0.2$), and high stellar probability (\texttt{CLASS\_STAR} $> 0.99$).

Figure \ref{fig:stellar locus} presents the resulting $(B-V)$ versus $(V-I)$ and $(V-I)$ versus $(V-R)$ color-color diagrams. The KS4 sources form a tight, well-defined stellar locus, indicating high internal photometric consistency. For comparison, we overplot reference loci derived from the \citet{1998PASP..110..863P} stellar spectral library for solar-metallicity dwarf (V) and giant (III) stars. The $(B-V)$ versus $(V-I)$ diagram (left panel) reveals a noticeable offset, highlighting the systematic deviation arising from the non-standard passband of the KMTNet $B$ filter relative to the standard Johnson-Cousins system {\refbf (e.g., \citealt{2017ApJ...848...19P, 2019ApJ...885...88P})}. Conversely, the $(V-I)$ versus $(V-R)$ diagram (right panel) demonstrates that the KS4 locus closely tracks the reference sequences, exhibiting only a slight redward deviation for the reddest K- and M-type stars.

Validating photometric accuracy for extended sources, such as galaxies, presents greater complexity than for point sources. This challenge is compounded by differences between the KS4 filter system and those employed by other deep surveys. Direct magnitude comparisons would necessitate complex color transformations that lie beyond the scope of this work. Therefore, we defer a detailed validation of extended source photometry to a forthcoming study focused on galaxy cluster searches using the red-sequence method (Park et al., in preparation).

\subsection{Catalog Validation}
\label{sec:catalog validation}
A direct photometric comparison between KS4 DR1 and other southern sky surveys is challenging due to fundamental differences in observational strategies, including disparate filter systems (e.g., Johnson-Cousins vs. Sloan-like), spatial coverage, and flux measurement methodologies. Consequently, we assess the quality of the KS4 catalog by quantifying the source recovery rate against external catalogs recognized for their high completeness and reliability: Gaia DR3 and DELVE DR2. Gaia DR3 provides a robust reference for bright sources, achieving nearly 100\% completeness for stars with $12 \lesssim G \lesssim 17$ and reliable coverage down to $G \approx 21$ \citep{2023A&A...674A...1G}. For fainter sources, the deeper DELVE DR2 catalog (median 5$\sigma$ depths of $g=24.3$ and $i=23.5$ mag) serves as a critical validation set \citep{2022ApJS..261...38D}. We note, however, that these references have limitations; Gaia completeness degrades in dense stellar fields, while the inhomogeneous footprint of DELVE restricts the available overlap area.

For this validation, we performed a source recovery analysis using representative KS4 tiles fully and homogeneously covered by both the Gaia DR3 and DELVE DR2 footprints. We selected two contrasting environments for this assessment: a dense, low Galactic latitude field (tile 0294; $b\approx -11^{\circ}$) with a source density of $\sim$54 sources arcmin$^{-2}$, and a sparse, high Galactic latitude field (tile 0698; $b\approx -82^{\circ}$) with a density of $\sim$10 sources arcmin$^{-2}$. We cross-matched our catalog with both reference surveys using a matching radius of 1.0 arcsec. Note that we did not apply proper motion corrections for the Gaia DR3 comparison. For the DELVE DR2 comparison, we restricted the reference sample to high-quality, non-spurious sources (\texttt{extended\_class} $\geq$ 0) detected in any of the $g$, $r$, $i$, or $z$ bands and observed in at least two epochs (\texttt{nepochs} $> 1$). Figure \ref{fig:completeness_recovery} presents the results of this analysis. In the sparse field, the KS4 catalog demonstrates high completeness, achieving nearly 100\% source recovery across most magnitude ranges relative to both Gaia DR3 (top panel) and DELVE DR2 (bottom panel). The comparison with DELVE also clearly illustrates the expected decline in completeness as the survey approaches its faint limit. In the dense field, however, the recovery rate is reduced; while bright sources maintain a recovery rate near 100\%, the overall completeness is lower compared to the sparse field. We attribute this degradation primarily to two factors prevalent in crowded regions: source confusion due to blending and the increased impact of detector bleeding from bright stars. The latter factor significantly contributes to the observed bright-end incompleteness by distorting source centroids, which causes objects located along vertical bleed trails to be missed (see Figure 4 in \citealt{2026arXiv2603.17442J}).

Beyond source recovery statistics, a distinct advantage of KS4 DR1 is its uniform and contiguous spatial coverage. Figure \ref{fig:completeness_distribution} compares a representative KS4 tile against seven other public surveys, highlighting the complementary value of this dataset. As illustrated, deep optical surveys such as DELVE DR2 and LS DR10 frequently exhibit irregular footprints and detector gaps. Similarly, infrared surveys like ALLWISE and VHS DR6 are often affected by saturation artifacts or incomplete tiling. In contrast, while Gaia DR3 and SMSS DR4 offer comparable spatial uniformity, KS4 DR1 extends to fainter magnitudes. Therefore, the KS4 catalog effectively complements existing datasets by providing continuous coverage in regions that are sparsely sampled or omitted by other major deep surveys.

\begin{figure}[!h]
    \centering
    \includegraphics[width=\linewidth]{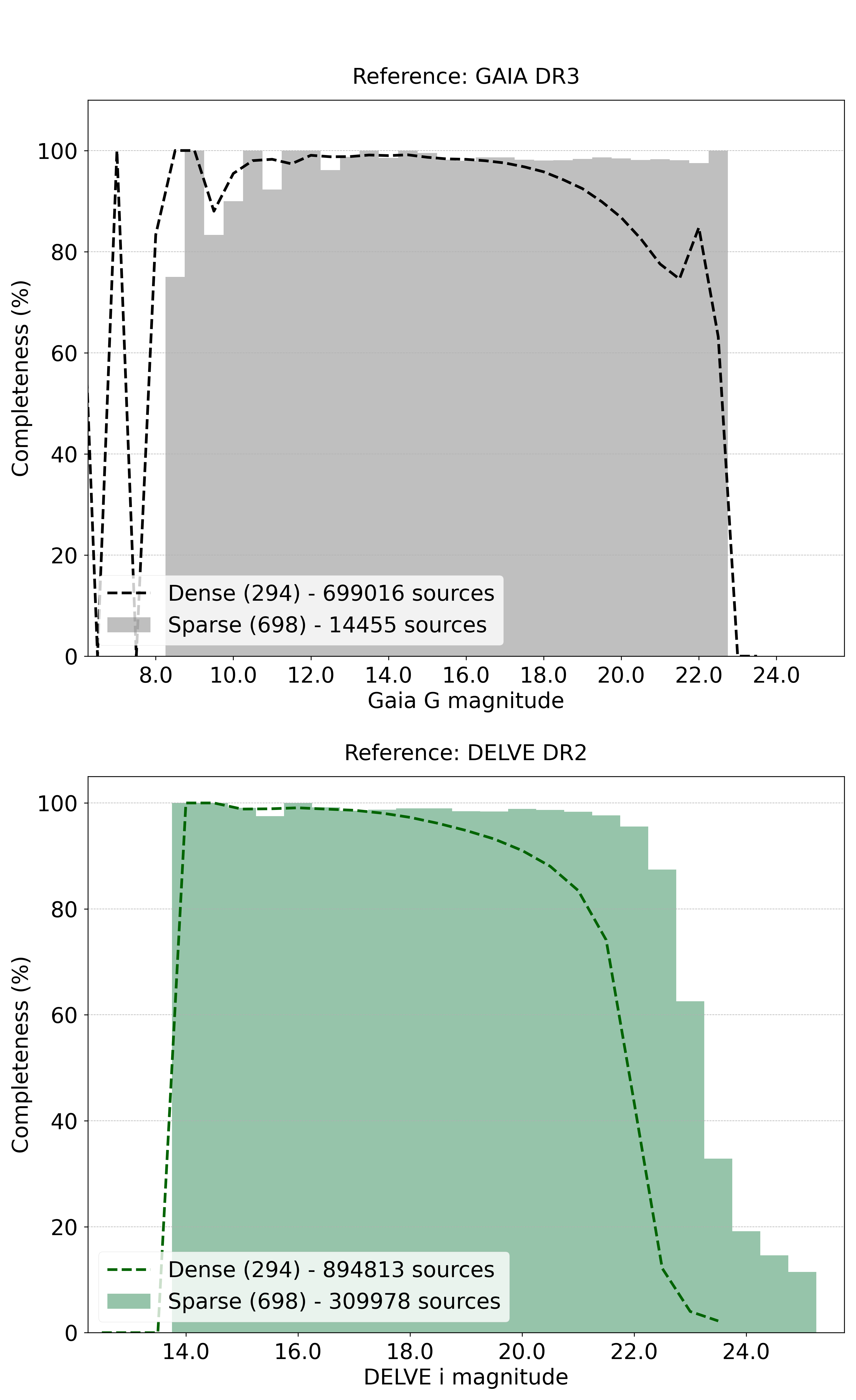}
    \caption{Source recovery completeness of the KS4 DR1 catalog, comparing a dense field (tile 0294, dashed line) and a sparse field (tile 0698, filled histogram). The completeness is shown relative to Gaia DR3 (top panel) and DELVE DR2 (bottom panel).}
   \label{fig:completeness_recovery}
\end{figure}

\begin{figure*}
    \centering
    \includegraphics[width=\linewidth]{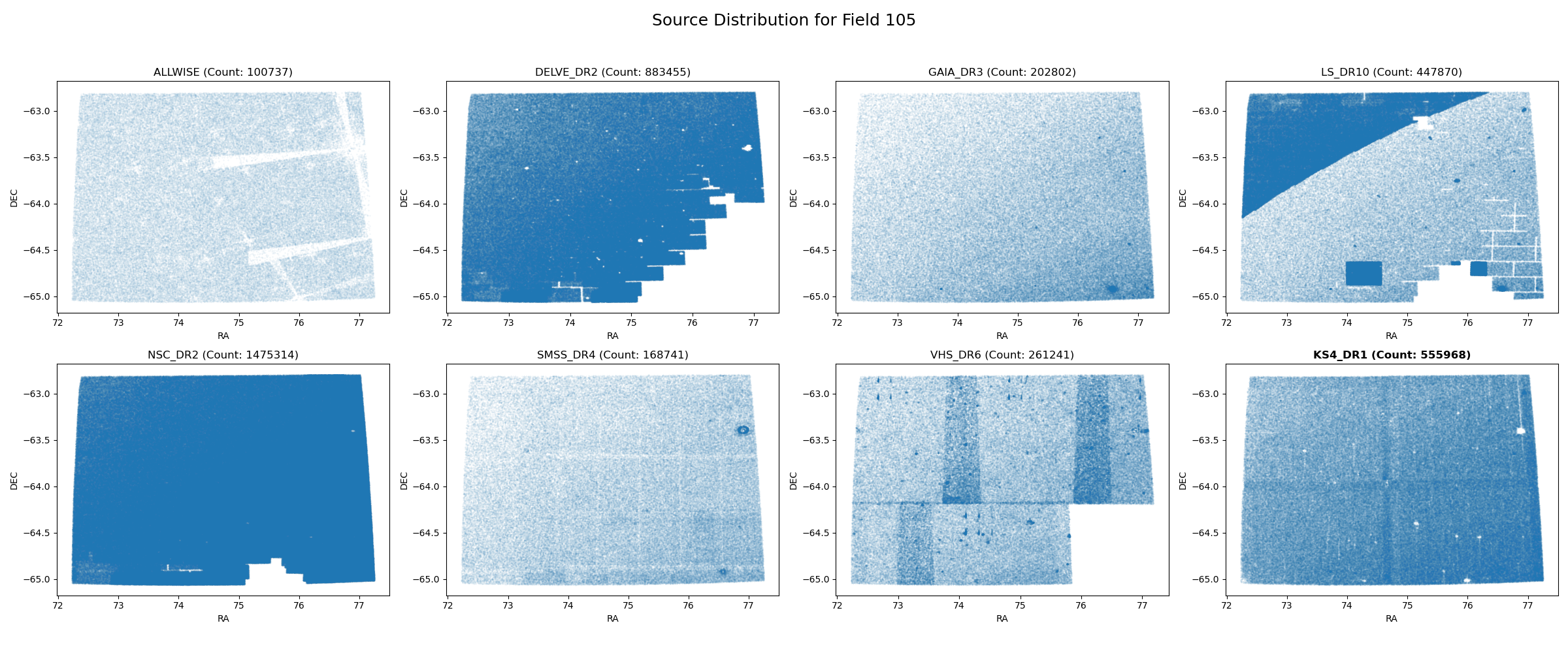}
    \caption{Comparison of the source spatial distributions from KS4 DR1 (bottom right) and seven other OIR surveys that have previously observed the same region of the sky, shown for an example KS4 tile 0105. The empty white regions in each panel correspond to areas that are either not covered by the footprint of each survey or where sources were removed by the corresponding data processing pipeline.}
  \label{fig:completeness_distribution}
\end{figure*}

\subsection{Known Limitations and Caveats}
\label{sec:known limitations}
This section outlines the primary limitations and caveats associated with the DR1 data products. Detailed descriptions of each issue are provided below. For visual examples of specific artifacts, such as flat-fielding residuals and stray light, we refer readers to Figure 12 of the companion pipeline paper \citep{2026arXiv2603.17442J}.

\begin{itemize}
    \item Flat-Fielding Residuals in High-Background Images: A subset of DR1 images exhibits vertical striping patterns along detector amplifier boundaries. We identify this artifact as a residual flat-fielding issue, which becomes prominent in images with high background levels caused by moonlight or cloud cover. The effect is most pronounced in the $I$-band and can introduce systematic biases in photometry, particularly for extended sources. We estimate this issue affects approximately 3\% (32 out of 979 fields) of the DR1 dataset, with a higher incidence in CTIO data obtained during July–August 2023. As specific flagging was not implemented for this release, we provide a list of affected fields in the public data repository (see Section 5.4 in \citealt{2026arXiv2603.17442J}).

    \item Stray Light Contamination: Stray light from bright stars (typically $V < 6$ mag) located outside the nominal field of view contaminates a fraction of the images. These artifacts manifest as faint streaks or ghost features. Their positions vary with telescope pointing, confirming their origin within the optical system. Due to their complex and variable morphology, systematic masking is challenging, potentially affecting source detection and photometry in contaminated regions. While our dithering and median-stacking strategy suppresses the majority of these artifacts, faint residuals may persist in the final co-added images.
    
    \item Residual Crosstalk: Standard pre-processing applies crosstalk correction using a master coefficient. However, temporal variations in this coefficient can lead to insufficient correction in frames with high source density (e.g., Galactic bulge fields) or elevated background levels. In such cases, standard algorithms fail to derive reliable per-image coefficients, leaving residual crosstalk artifacts that may mimic transient sources in difference imaging. We attempted to predict locations of victim pixels by tracing saturated stars (peak flux $> 56,000$ ADU); however, users should note that this predictive flagging is not exhaustive (see Section 3.3.2 in \citealt{2026arXiv2603.17442J}). 

   \item {\refbf Uncorrected Saturated Sources: Saturated pixels in individual exposures are flagged (\texttt{FLAGS=4}) and propagated to the final catalogs via the \texttt{IMAFLAGS\_ISO} parameter \citep{2026arXiv2603.17442J}. However,  the ZP scaling process alters pixel values, resulting in inconsistent saturation thresholds in the stacked images. Furthermore, saturation levels (44,000--55,000 ADU) and bleeding directions vary significantly across CCD chips. While photometric correction is feasible through comparison with external catalogs (e.g., Gaia XP), its robust implementation across the entire dataset was deferred to future releases.}


    \item Systematic Photometric Offsets near Faulty Amplifiers: Comparison with Gaia XP synthetic photometry reveals systematic offsets in detector regions affected by faulty amplifiers, where KS4 magnitudes can be measured up to 1 mag fainter than expected. These deviations are most pronounced near amplifier boundaries, where incomplete detector coverage leads to poorly constrained photometry. Although we attempted to correct for this effect using spatially coherent source groups, the method proved unreliable in regions with a low density of Gaia XP reference stars. Thus, no correction has been applied in DR1. Users requiring high precision should exclude sources where the 8-bit flag (indicating a faulty readout port) is set in \texttt{IMAFLAGS\_ISO}.

    \item Site-Dependent Color Terms:  A primary limitation in DR1 photometry results from the co-addition of images from the three KMTNet sites without correcting for site-specific filter transmission variations. Despite identical manufacturing specifications, physical differences in passbands are non-negligible. The variation is most pronounced in the $V$-band, which has a relatively narrow passband compared to the broader, less sensitive $I$-band. For instance, the $V$-filter passband at CTIO spans 513.79–596.7 nm, differing notably from the ranges at SAAO (507.09–593.21 nm) and SSO (507.12–592.59 nm). These physical differences result in a central wavelength variation of more than 5 nm between sites. These variations introduce significant site-dependent color terms relative to Gaia XP. Since images from different sites are often combined, these uncorrected differences can introduce systematic biases in the final catalog photometry, particularly for objects with very red or blue spectral energy distributions. This effect is most prominent in co-added images that include data from the CTIO site. {\refbf Because DR1 co-added images combine exposures from multiple sites without applying site-specific photometric corrections, we do not assign site-level flags to the final catalogs.}
   
\end{itemize}

\begin{figure*}
    \centering
    \includegraphics[width=\linewidth]{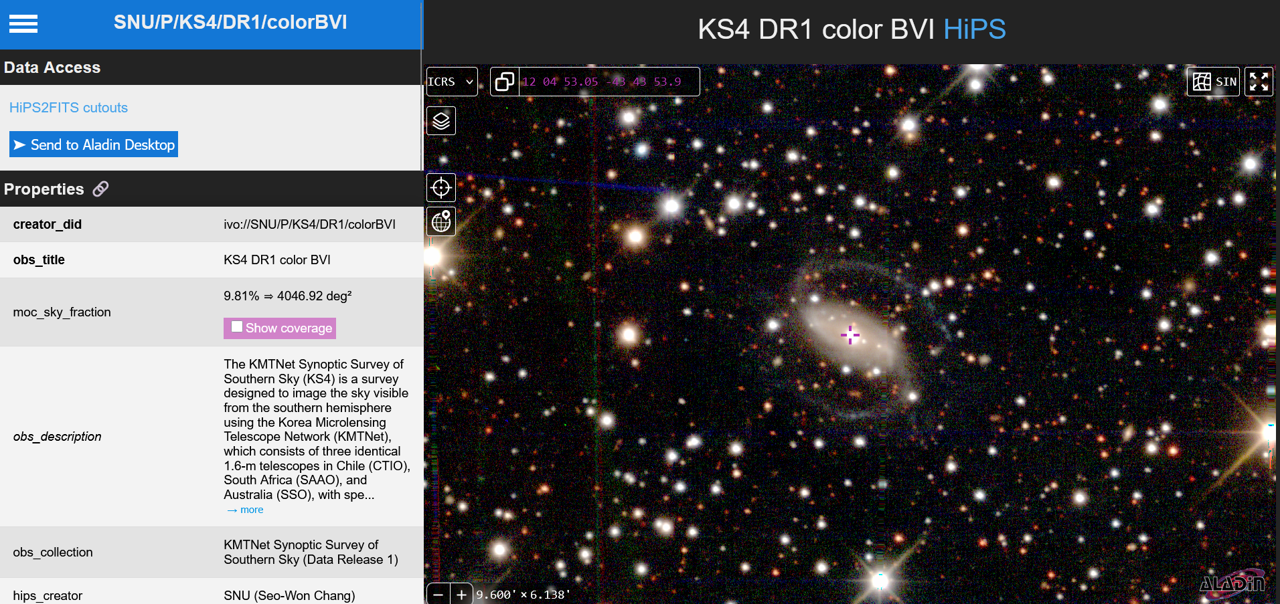}
    \caption{The landing page for the KS4 DR1 three-color ($BVI$) composite HiPS dataset, as displayed in the Aladin Lite web interface. The screenshot shows the default view. The metadata panel shown on the left provides key information about the survey and direct links to services such as \texttt{hips2fits} cutouts.}
    \label{fig:KS4 DR1 HiPS}
\end{figure*}


\section{Data Access and Distribution} 
\label{sec:data access}
To ensure broad and sustainable access for the astronomical community, KS4 DR1 data products are distributed via a multi-platform strategy. We host data through established international data centers to guarantee stable, long-term availability. Imaging data are accessible through both interactive visualization platforms and direct download servers, while source catalogs are hosted on large-scale, query-optimized databases.

\subsection{Images}
\label{sec:image access}
For interactive visualization and rapid data exploration, we provide imaging data in the HiPS format, registered in the CDS HiPS registry\footnote{\url{https://alasky.cds.unistra.fr/hipslist}}. A web-based visualization of the KS4 DR1 HiPS datasets is available at \url{https://alasky.cds.unistra.fr/KS4/DR1/}. Direct access points for the individual datasets are:
\begin{itemize}
\item Color ($BVI$): \url{https://alasky.cds.unistra.fr/KS4/DR1/SNU_P_KS4_DR1_colorBVI/}
\item $B$-band: \url{https://alasky.cds.unistra.fr/KS4/DR1/SNU_P_KS4_DR1_B/}
\item $V$-band: \url{https://alasky.cds.unistra.fr/KS4/DR1/SNU_P_KS4_DR1_V/}
\item $R$-band: \url{https://alasky.cds.unistra.fr/KS4/DR1/SNU_P_KS4_DR1_R/}
\item $I$-band: \url{https://alasky.cds.unistra.fr/KS4/DR1/SNU_P_KS4_DR1_I/}
\end{itemize} These resources enable seamless visualization and browsing of KS4 DR1 images within standard astronomical software, such as Aladin, ESASky, and the ESA Science Portal. Figure \ref{fig:KS4 DR1 HiPS} presents a screenshot of the three-color composite HiPS, illustrating the interactive exploration capabilities of these tools. For researchers requiring FITS-format cutouts of specific regions, the CDS \texttt{hips2fits}\footnote{\url{https://alasky.cds.unistra.fr/hips-image-services/hips2fits}} service allows the generation of cutouts with arbitrary size and resolution. This service is programmatically accessible for scripted workflows via the \texttt{astroquery} Python package \citep{2019AJ....157...98G}.

Bulk downloads of full-scale image products—including deep co-added images, single-epoch exposures, and associated masks—are currently available via direct transfer from our institutional server upon request. {\refbf For typical use cases involving specific targets, we recommend the \texttt{hips2fits} service for its efficiency in providing customized image cutouts.}


\subsection{Catalogs}
\label{sec:catalog access}
The primary repository for both the \texttt{idual\_master} and \texttt{single\_master} source catalogs is the Astro Data Lab \citep{2014SPIE.9149E..1TF,2020A&C....3300411N} at National Science Foundation's NOIRLab\footnote{\url{https://datalab.noirlab.edu/data-explorer}}. Hosted within the External Surveys collection, these catalogs utilize the platform's substantial storage resources to enable complex, server-side data analysis. 

Upon publication, the band-merged (\texttt{single\_master}) catalog will also be accessible via the VizieR service\footnote{\url{https://vizier.cds.unistra.fr/}} at CDS. This distribution will include a HiPS-compatible version of the catalog, enabling interactive visualization and source selection within Aladin and other HiPS-enabled clients.

To support multi-wavelength studies, the catalogs hosted on Astro Data Lab will be cross-matched against standard external datasets, such as Gaia DR3, NSC DR2, unWISE, ALLWISE, DELVE DR2, DELVE DR3 (Drlica-Wagner et al., in prep), and SMSS DR4. Additionally, we can provide our own pre-matched tables generated for internal analysis upon request, although these are not part of the formal data release.

{\refbf To support the effective use of the DR1 catalogs, we provide Jupyter notebook tutorials via the Astro Data Lab platform. These notebooks illustrate recommended quality-selection criteria and demonstrate data access methods for common science applications, including guidance on interpreting key quality parameters (e.g., \texttt{FLAGS}, \texttt{IMAFLAGS\_ISO}, and \texttt{NDITH}) and example SQL queries for pre-matched external catalogs\footnote{\url{https://datalab.noirlab.edu/data/ks4}}.}

\subsection{Zero-Point Maps}
\label{sec:zp map access}
The 2D correction maps detailed in Section \ref{sec:zp correction map} are archived on Zenodo (\href{https://doi.org/10.5281/zenodo.17336382}{Jeong 2025})\footnote{\url{https://zenodo.org/records/17336382}}. For each combination of KS4 tile, filter, and detection mode, we provide data products in NumPy array format. These include the ZP correction map (\texttt{zp\_map}) and its associated uncertainty (\texttt{zperr\_map}), as well as the astrometric alignment offset map (\texttt{align\_map}) and its uncertainty (\texttt{alerr\_map}).

To simplify the use of these maps, we provide a Python toolkit hosted on a public GitHub repository\footnote{\url{https://github.com/jmk5040/KMTNet_ToO/blob/main/pipe/}}. This script includes helper functions that enable users to query correction values for individual targets based on pixel coordinates (X, Y) or to efficiently apply corrections to entire source catalogs.

\section{Summary and Future Prospects}
\label{sec:summary}
We present the first public data release, DR1, from the KMTNet Synoptic Survey of the Southern Sky. This release provides science-ready data products for approximately $4,000~\rm{deg}^{2}$ of the southern sky in the $B$, $V$, $R$, and $I$ bands, reaching median $5\sigma$ limiting magnitudes of 22.0--23.5 mag. The data products demonstrate high astrometric precision, characterized by a modal offset of 0.125 arcsec relative to Gaia DR3. Furthermore, the photometry exhibits a high degree of uniformity, with median offsets relative to Gaia XP remaining within $\pm0.03$ mag across the majority of the survey footprint. The primary value of this release lies in its uniform and contiguous spatial coverage. The KS4 dataset extends to fainter magnitudes than other uniform surveys (e.g., Gaia, SMSS DR4) while filling the irregular gaps present in deep surveys such as DELVE DR2 and LS DR10. DR1 comprises deep co-added images, ancillary zero-point correction maps, and two distinct photometric source catalogs containing over 279 million unique entries, including over 200 million high-confidence sources with SNR $>5$.

To ensure broad and sustainable access for the astronomical community, we distribute KS4 DR1 data products through a multi-platform strategy hosted by established international data centers. The two source catalogs, \texttt{idual\_master} and \texttt{single\_master}, are available via query-optimized databases at NOIRLab's Astro Data Lab and the VizieR service at CDS. The imaging data can be explored interactively via HiPS, with full-resolution FITS cutouts available through the \texttt{hips2fits} service, providing robust, long-term access for a wide range of scientific use cases.

This work marks the first in a series of planned releases. For the second data release (DR2), we plan to expand coverage to the entire KS4 survey area observed to date, as shown in Figure \ref{fig:coverage_map}, increasing the total footprint to over 6,700 $\rm{deg}^{2}$. Furthermore, DR2 will include photometric measurements from individual single-epoch images, enabling comprehensive time-domain analyses. Future releases will also incorporate ongoing improvements to the data reduction and calibration pipelines to further enhance the quality and reliability of the KS4 data products.


\acknowledgments
We begin by acknowledging the funding sources that made this research possible. This work was supported by the National Research Foundation of Korea (NRF) grants, No. 2021R1A2C3009648 and No. 2021M3F7A1084525, funded by the Ministry of Science and ICT (MSIT). SWC acknowledges support from the NRF grants funded by the Ministry of Education (RS-2023-00245013) and the MIST (RS-2026-25489059). This research was also funded in part by the Australian Research Council Centre of Excellence for Gravitational Wave Discovery (OzGrav), CE170100004 and CE230100016. JC acknowledges funding from the Australian Research Council Discovery Project DP200102102. G. S. H. P. acknowledges support from the Pan-STARRS project, which is a project of the Institute for Astronomy of the University of Hawai'i, and is supported by the NASA SSO Near Earth Observation Program under grants 80NSSC18K0971, NNX14AM74G, NNX12AR65G, NNX13AQ47G, NNX08AR22G, 80NSSC21K1572, and by the State of Hawai'i. Y. K. was supported by the faculty research fund of Sejong University in 2025 and the NRF grant funded by the Korean government (MSIT) (No. 2021R1C1C2091550). SL and BP acknowledge support from the NRF grant (RS-2025-00573214) funded by the Korea government (MSIT).

This research has made use of the KMTNet system operated by the Korea Astronomy and Space Science Institute (KASI) at three host sites of CTIO in Chile, SAAO in South Africa, and SSO in Australia.
Data transfer from the host site to KASI was supported by the Korea Research Environment Open NETwork (KREONET).

We are grateful to the teams at the international data centers for their essential support in hosting and distributing the KS4 DR1 data products. We thank the team from the Astro Data Lab, a program of NSF's NOIRLab, particularly Brian Merino and Robert Nikutta, for helping make DR1 available on their platform. We also thank Thomas Boch from the Centre de Données astronomiques de Strasbourg (CDS) for their support in the creation and distribution of our HiPS datasets and for making the catalog available via the VizieR service. This research uses services or data provided by the Astro Data Lab, which is part of the Community Science and Data Center (CSDC) Program of NSF NOIRLab. NOIRLab is operated by the Association of Universities for Research in Astronomy (AURA), Inc. under a cooperative agreement with the U.S. National Science Foundation. Furthermore, we thank the Globus tool \citep{2011IEEE..15..70, 2012CACM..55..81} for providing a fast and reliable service that facilitated our large-scale data transfers.

This work has made use of data from the European Space Agency (ESA) mission {\it Gaia} (\url{https://www.cosmos.esa.int/gaia}), processed by the {\it Gaia} Data Processing and Analysis Consortium (DPAC, \url{https://www.cosmos.esa.int/web/gaia/dpac/consortium}). Funding for the DPAC has been provided by national institutions, in particular the institutions participating in the {\it Gaia} Multilateral Agreement.

This research has made extensive use of TOPCAT: Tool for OPerations on Catalogues And Tables, developed by Mark Taylor, for data analysis and figure generation \citep{2005ASPC..347...29T}. This research made use of Astropy, a community-developed core Python package for Astronomy \citep{2013A&A...558A..33A, 2018AJ....156..123A}.


\appendix
\section{KS4 DR1 Catalog Schema}
\label{sec:appendix_schema}
Table \ref{tab:dr1 schema} details the column definitions common to both the \texttt{idual\_master} and \texttt{single\_master} catalogs. In the column nomenclature, the placeholder \{band\} denotes one of the four photometric filters ($B, V, R, I$), while \{n\} corresponds to the fixed aperture radii of 3, 5, or 10 arcseconds. 

The catalogs incorporate several bitwise integer flags for quality assessment. For the primary status flags, \texttt{flags} and \texttt{imaflags\_iso}, a value of 0 signifies a clean source free of known issues. The \texttt{flags} column is an 8-bit integer containing the standard extraction warnings raised by SExtractor \citep{1996A&AS..117..393B}. The \texttt{imaflags\_iso} column captures information from the external pipeline mask (see \citealt{2026arXiv2603.17442J}) mapped to the source's isophotal footprint. 

Supplementary flags are provided to evaluate source reliability and observational coverage. The \texttt{ssflag} identifies potential spurious detections (see Section \ref{sec:removal of spurious sources}), while \texttt{sscount\_30} quantifies the local density of removed spurious sources within a 30 arcsec radius. Additionally, the \texttt{ndith} column records the number of contributing single-epoch exposures, and the \texttt{edge} flag indicates sources located in the proximity of tile boundaries.

\begin{table*} 
\centering
\caption{Name, description, and units for each column in the \texttt{idual\_master} and \texttt{single\_master} table.}
\label{tab:dr1 schema}
\begin{tabular}{p{0.25\textwidth}p{0.6\textwidth}p{0.1\textwidth}}
\hline
Column Name & Description & Units \\
\hline
coadd\_object\_id & Unique object identifier & -- \\
ra & Right Ascension (J2000) & deg \\
dec & Declination (J2000) & deg \\
healpix\_index & HEALPix index for sky position & -- \\
n\_det & Number of detections across all KS4 tiles & -- \\
glon & Galactic longitude & deg \\
glat & Galactic latitude & deg \\
ebmv\_sfd & E(B-V) extinction from \citet{1998ApJ...500..525S} using the \texttt{dustmap} tool & mag \\
mag\_auto\_\{band\} & SExtractor AUTO magnitude in \{band\} & mag \\
magerr\_auto\_\{band\} & Error in AUTO magnitude in \{band\} & mag \\
mag\_aper\{n\}\_\{band\} & Aperture magnitude (\{n\} arcsec) in \{band\} & mag \\
magerr\_aper\{n\}\_\{band\} & Error in aperture magnitude (\{n\} arcsec) in \{band\} & mag \\
min\_mjd\_\{band\} & Minimum Modified Julian Date in \{band\} & d \\
max\_mjd\_\{band\} & Maximum Modified Julian Date in \{band\} & d \\
mean\_mjd\_\{band\} & Mean Modified Julian Date in \{band\} & d \\
fwhm\_image\_\{band\} & FWHM of the object in \{band\} & arcsec \\
class\_star\_\{band\} & Source Extractor stellarity index in \{band\} & -- \\
flux\_radius\_\{band\} & Half-light radius in \{band\} & arcsec \\
kron\_radius\_\{band\} & Kron radius factor used for \texttt{MAG\_AUTO} in \{band\}  & -- \\
flux\_max\_\{band\} & Peak count above background of the brightest object pixel in \{band\} & counts \\
theta\_image\_\{band\} & Position angle in \{band\} & deg \\
ellipticity\_\{band\} & Ellipticity in \{band\} & -- \\
a\_image\_\{band\} & Length of object semi-major axis in pixels in \{band\} & pixel \\
b\_image\_\{band\} & Length of object semi-minor axis in pixels in \{band\} & pixel \\
flags\_\{band\} & Flags raised by Source Extractor in \{band\} & -- \\
imaflags\_iso\_\{band\} & OR-combined flagged pixel types within the isophotal radius in \{band\} & -- \\
ssflag\_\{band\} & Spurious source flag in \{band\} & -- \\
sscount\_30\_\{band\} & Spurious source count within 30 arcsec in \{band\} & -- \\
ndith\_\{band\} & Number of all contributing single-epoch exposures in \{band\} & -- \\
edge\_\{band\} & Edge flag in \{band\} & -- \\
\hline
\end{tabular}
\end{table*}


\end{document}